\documentclass[12pt]{iopart}
\usepackage[numbers,square,sort&compress]{natbib}
\usepackage{graphicx,bm}
\usepackage{hyperref}

\newcommand{\aap}{{Astr. \& Astrophys. \/}}

\newcommand{\apj}{{Astrophys. J. \/}}
\newcommand{\apjl}{{Astrophys. J. Lett. \/}}
\newcommand{\apjs}{{Astrophys. J. Suppl. \/}}

\newcommand{\mnras}{{Mon. Not. R. Astr. Soc. \/}}
\newcommand{\prd}{{Phys. Rev. D\/}}
\newcommand{\prl}{{Phys. Rev. Lett. \/}}
\newcommand{\physrep}{{\em Phys. Rep. \/}}
\newcommand{\fnlKS}  {f_{\rm NL}^{\rm local}}
\newcommand{\fnleq}  {f_{\rm NL}^{\rm equil}} 
\newcommand{\fnlor}  {f_{\rm NL}^{\rm orthog}} 
\newcommand{\bsrc} {b_{\rm src}}
\begin{document}
\title{Hunting for primordial non-Gaussianity in the cosmic microwave
background}
\author{Eiichiro Komatsu}
\address{Texas Cosmology Center and Department of Astronomy, The
University of Texas at Austin, Austin, TX 78712, USA}
\ead{komatsu@astro.as.utexas.edu}
\begin{abstract}
 Since the first limit on the (local) primordial non-Gaussianity parameter,
 $f_{\rm NL}$, was obtained from the {\sl Cosmic Background Explorer}
 ({\sl COBE}) 
 data in 2002, observations of the cosmic microwave background (CMB)
 have been playing a central role in constraining the amplitudes of 
 various forms of non-Gaussianity in primordial fluctuations. The
 current 68\% limit from the 7-year
 data of the {\sl Wilkinson Microwave Anisotropy Probe} ({\sl WMAP})
 is $f_{\rm NL}=32\pm 21$, and the {\sl Planck} satellite is expected to
 reduce the uncertainty by a factor of four in a few years from now. If
 $f_{\rm NL}\gg 1$ is found by {\sl Planck} with high statistical
 significance, all single-field 
 models of inflation would be ruled out. Moreover, if the {\sl Planck}
 satellite 
 finds $f_{\rm NL}\sim 30$, then it would be able to test a broad class of
 multi-field models using the four-point function (trispectrum) test of
 $\tau_{\rm 
 NL}\ge (6f_{\rm NL}/5)^2$.
 In this article, we review the methods (optimal estimator), results
 ({\sl WMAP} 7-year), and challenges (secondary anisotropy, second-order
 effect, and foreground) of
 measuring primordial non-Gaussianity from the CMB data, present a
 science case for the trispectrum, and conclude with
 future prospects.
\end{abstract}
\section{Introduction}
The physics of the very early, primordial universe is best probed by
measurements of statistical properties of primordial fluctuations. 
The primordial fluctuations are
the seeds for the temperature and polarization anisotropies of the CMB
and the large-scale structure of the universe that we observe
today. Therefore, both the CMB and the large-scale structure are
excellent probes of the primordial fluctuations. In 
this article, 
we shall focus on the CMB. See the article by V. Desjacques and
U. Seljak in this volume for the 
corresponding review on 
the large-scale structure as a probe of the primordial fluctuations.

This article reviews a recent progress on our using the CMB as a probe
of a particular statistical aspect of primordial fluctuations called
``{\it non-Gaussianity}.''  
Reviews on this subject were written in 2001 \cite{komatsu:prep} and
2004 \cite{bartolo/etal:2004}. The former review would be most useful
for those who are new to this subject.

In this article, we focus on the new discoveries that have been made since 2004.
Particularly notable ones include:
\begin{itemize}
 \item[1.] It has been proven that {\it all} inflation models (not just
	   simple ones 
	   \cite{maldacena:2003,acquaviva/etal:2003}) based upon 
	   a single scalar field would be ruled out regardless of the
	   details of models \cite{creminelli/zaldarriaga:2004}, if the
	   primordial non-Gaussianity 
	   parameter called $f_{\rm NL}$ (more precisely, the 
	   ``local type'' $f_{\rm NL}$ as described later) is
	   found to be much greater than unity.
 \item[2.] The optimal method for extracting $f_{\rm NL}$ from the CMB data has
	   been developed
	   \cite{komatsu/spergel/wandelt:2005,creminelli/etal:2006,smith/zaldarriaga:prep,yadav/etal:2008}
	   and implemented \cite{smith/senatore/zaldarriaga:2009}. The
	   latest limit on the local-type $f_{\rm NL}$ from the
	   {\sl WMAP} 7-year temperature data is $f_{\rm NL}=32\pm 21$
	   (68\%~CL) \cite{komatsu/etal:prep}.
 \item[3.] The most serious contamination of the local-type $f_{\rm
	   NL}$ due to the secondary CMB anisotropy, the coupling
	   between the Integrated Sachs-Wolfe (ISW) 
	   effect and the weak gravitational lensing, has been
	   identified
	   \cite{goldberg/spergel:1999,verde/spergel:2002,smith/zaldarriaga:prep,serra/cooray:2008,hanson/etal:2009,mangilli/verde:2009}. However,
	   note that the astrophysical contamination such as the
	   Galactic foreground emission and radio point sources may
	   still be the most serious contaminant of $f_{\rm NL}$. These
	   effects would pose a serious analysis challenge to measuring
	   $f_{\rm NL}$ from the {\sl Planck} data.
 \item[4.] The importance of distinguishing different triangle
	   configurations of the three-point function of the CMB was
	   realized
	   \cite{creminelli:2003,babich/creminelli/zaldarriaga:2004} and
	   has been fully appreciated. It has
	   been shown by many researchers that different configurations
	   probe distinctly 
	   different aspects of the physics of the primordial
	   universe. The list of possibilities is long, and a terribly
	   incomplete list of references on recent work (since $\sim$
	   2004) is: \cite{chen/etal:2007} on a general analysis of
	   various shapes; 
	   \cite{lyth/ungarelli/wands:2003,zaldarriaga:2004,lyth/rodriguez:2005,lehners:2008,bond/etal:2009,chambers/nurmi/rajantie:2010}
	   on the local shape ($k_3\ll k_1\approx k_2$); \cite{creminelli:2003,alishahiha/silverstein/tong:2004,arkani-hamed/etal:2004,seery/lidsey:2005} on the
	   equilateral shape ($k_1\approx k_2\approx k_3$);
	   \cite{holman/tolley:2008,meerburg/etal:2009} on the flattend
	   (or folded) shape 
	   ($k_1\approx 2k_2\approx 2k_3$);
	   \cite{senatore/smith/zaldarriaga:2010,huang:prep} on the orthogonal
	   shape (which is nearly orthogonal to both local and
	   equilateral shapes);  
	   \cite{moss/chun:2007,arroja/mizuno/koyama:2008,renaux-petel:2009,meerburg/etal:2010,chen/wang:prep} 
	   on 
	   combinations of different shapes; and
	   \cite{chen/easther/lim:2007,chen/easther/lim:2008}
	   on 
	   oscillating bispectra. Also 
	   see references therein.
 \item[5.] The connected four-point function of primordial fluctuations has been
	   shown to be an equally powerful probe of the physics of the
	   primordial universe. In particular, a combination of the
	   three- and four-point functions may allow us to further
	   distinguish different scenarios. Many papers have been
	   written on this subject over the last few years: 
 \cite{arroja/koyama:2008,arroja/etal:2009,chen/etal:2009,seery/sloth/vernizzi:2009} 
	   on single-field models; 
	   \cite{boubekeur/lyth:2006,huang/shiu:2006,byrnes/sasaki/wands:2006,seery/lidsey:2007,seery/lidsey/sloth:2007,suyama/yamaguchi:2008,suyama/takahashi:2008,ichikawa/etal:2008,ichikawa/etal:2008b,cogollo/rodriguez/valenzuela-toledo:2008,rodriguez/valenzuela-toledo:2008,buchbinder/khoury/ovrut:2008,mizuno/etal:2009,mizuno/arroja/koyama:2009,gao/li/lin:2009,gao/hu:2009,byrnes/choi/hall:2009,byrnes/tasinato:2009,enqvist/takahashi:2008,enqvist/etal:2009,enqvist/takahashi:2009,kawasaki/takahashi/yokoyama:2009,renaux-petel:2009,huang:2008,huang:2009,chingangbam/huang:2009,chen/wang:prepb}
	   on multi-field models; and
	   \cite{kawakami/etal:2009,takahashi/yamaguchi/yokoyama:2009} on isocurvature perturbations.
	   CMB data are expected to provide useful limits on the
	   parameters of the ``local-form trispectrum,'' 
	   $\tau_{\rm NL}$ and $g_{\rm NL}$
	   \cite{okamoto/hu:2002,kogo/komatsu:2006}. Preliminary limits
	   on these parameters have been obtained from the {\sl WMAP} data by
	   \cite{vielva/sanz:prep,smidt/etal:prep}. 
\end{itemize}
The number of researchers working on primordial non-Gaussianity
has increased dramatically: Science White Paper on non-Gaussianity
submitted to Decadal Survey Astro2010 was co-signed by 61 scientists
\cite{komatsu/etal:astro2010}.

\section{Gaussian versus non-Gaussian CMB anisotropy}
\subsection{What do we mean by ``Gaussianity''?}
What do we mean by ``Gaussian fluctuations''?
Let us consider the distribution of temperature anisotropy of the CMB that we
observe on the sky, $\Delta T(\hat{\bm{n}})$. The temperature anisotropy
is Gaussian when its probability density function (PDF) is given by 
\begin{equation}
 P(\Delta T)=
\frac1{(2\pi)^{N_{\rm pix}/2}|\xi|^{1/2}}
\exp\left[-\frac12\sum_{ij} \Delta T_i(\xi^{-1})_{ij}\Delta
		  T_j \right],
\label{eq:gaus1}
\end{equation}
where $\Delta T_i\equiv \Delta T(\hat{\bm{n}})$, $\xi_{ij}\equiv
\langle \Delta T_i\Delta T_j\rangle$ is the covariance matrix (or the
two-point correlation function) of the temperature anisotropy, $|\xi|$
is the determinant of the covariance matrix, and $N_{\rm pix}$ is the
number of pixels on the sky.

\begin{figure}[t]
\centering \noindent
\includegraphics[width=14cm]{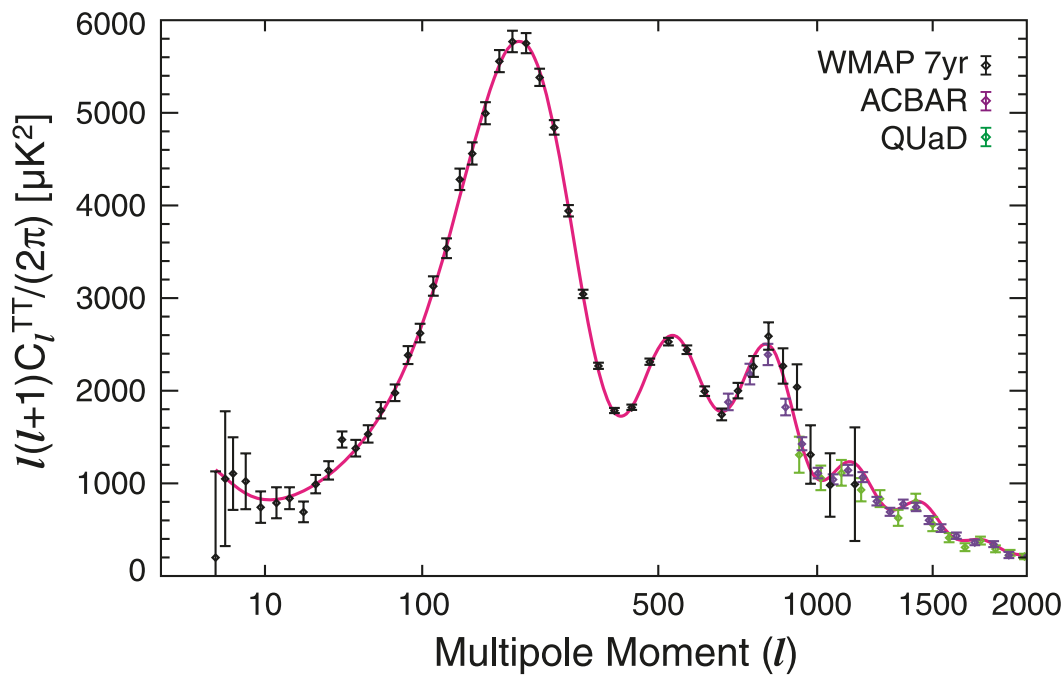}
\caption{%
 The angular power spectrum of the CMB temperature anisotropy, $C_l$,
 measured from the {\sl WMAP} 7-year data \cite{larson/etal:prep}, 
 along with the temperature power spectra from the ACBAR
 \cite{reichardt/etal:2009} and QUaD \cite{brown/etal:2009}
 experiments. The solid line shows the best-fitting 6-parameter flat
 $\Lambda$CDM model to the {\sl WMAP} data alone.  The angular power
 spectrum contains all 
 the information on fluctuations in the CMB, if fluctuations are
 Gaussian. If fluctuations are non-Gaussian, one must use the
 higher-order correlation functions (such as three- and four-point
 functions) to fully exploit the cosmological information contained in
 the CMB.
 This figure is adopted
 from \cite{komatsu/etal:prep}.
} 
\label{fig:cl}
\end{figure}

We often work in harmonic space by expanding $\Delta T$ using spherical
harmonics: $\Delta
T(\hat{\bm{n}})=\sum_{lm}a_{lm}Y_{lm}(\hat{\bm{n}})$. The PDF for $a_{lm}$ is given by
\begin{equation}
 P(a)=
\frac1{(2\pi)^{N_{\rm harm}/2}|C|^{1/2}}
\exp\left[-\frac12\sum_{lm}\sum_{l'm'}a^*_{lm}(C^{-1})_{lm,l'm'}a_{l'm'}\right],
\label{eq:gaus2}
\end{equation}
where $C_{lm,l'm'}\equiv \langle a^*_{lm}a_{l'm'}\rangle$, and $N_{\rm
harm}$ is the number of $l$ and $m$.
When $a_{lm}$ is statistically homogeneous and isotropic (which is not
always the case 
because of, e.g., non-uniform noise), one finds
$C_{lm,l'm'}=C_l\delta_{ll'}\delta_{mm'}$, and thus the PDF simplifies to
\begin{equation}
 P(a)=\prod_{lm}\frac{e^{-|a_{lm}|^2/(2C_l)}}{\sqrt{2\pi C_l}}.
\label{eq:gaus3}
\end{equation}
 Here, $C_l$ is
the {\it angular power spectrum}. The latest determination of $C_l$ of the CMB
temperature anisotropy is shown in Figure~\ref{fig:cl}.

The important property of a Gaussian distribution is that the
PDF is fully specified by the covariance matrix. In
other words, the covariance matrix contains all the information on
statistical properties of Gaussian fluctuations. When the PDF is given
by equation~\ref{eq:gaus3}, the power spectrum, 
$C_l$, contains all the information on $a_{lm}$. This is not true for
non-Gaussian fluctuations, for which one needs information on
higher-order correlation functions.

Let us close this subsection by noting that a non-zero deviation of the
covariance matrix from the diagonal form, $\Delta
C_{lm,l'm'}=C_{lm,l'm'}-C_l\delta_{ll'}\delta_{mm'}$, does not imply 
 non-Gaussianity: the PDF can be a Gaussian with
a non-diagonal covariance matrix as given in equation~\ref{eq:gaus2}. A
non-zero $\Delta C_{lm,l'm'}$ may arise in cosmological models that
violate statistical isotropy. Such models may yield {\it anisotropic Gaussian}
fluctuations; thus, one must distinguish between non-Gaussianity and
a violation of statistical isotropy.

\subsection{What do we mean by ``non-Gaussianity''?}
What do we mean by ``non-Gaussian fluctuations''?
Any deviation from a Gaussian distribution (such as
equation~\ref{eq:gaus1} or \ref{eq:gaus2})
is called {\it non-Gaussianity}.
When fluctuations in the CMB are non-Gaussian, one cannot generally
write down its PDF, 
unless one considers certain models (e.g., 
inflation). Nevertheless, when non-Gaussianity is weak, one may 
expand the PDF around a Gaussian distribution \cite{taylor/watts:2001}
and obtain
\begin{eqnarray}
\nonumber
 P(a)&=&
\left[1-\frac16\sum_{{\rm all}~l_im_j}
\langle a_{l_1m_1}a_{l_2m_2}a_{l_3m_3}\rangle
\frac{\partial}{\partial a_{l_1m_1}}
\frac{\partial}{\partial a_{l_2m_2}}
\frac{\partial}{\partial a_{l_3m_3}}
\right]\\
& &\times \frac{e^{-\frac12\sum_{lm}\sum_{l'm'}a^*_{lm}(C^{-1})_{lm,l'm'}a_{l'm'}}}{(2\pi)^{N_{\rm harm}/2}|C|^{1/2}}.
\end{eqnarray}
Here, the expansion is truncated at the three-point function
(bispectrum) of $a_{lm}$, and thus we have assumed that the connected four-point
and higher-order
correlation 
functions are negligible compared to the power spectrum and
bispectrum. (This condition is not always satisfied.)
By evaluating the above derivatives, one obtains\footnote{Babich
\cite{babich:2005} derived this formula for
$C_{lm,l'm'}=C_l\delta_{ll'}\delta_{mm'}$.} 
\begin{eqnarray}
\nonumber
& &P(a)=
\frac1{(2\pi)^{N_{\rm harm}/2}|C|^{1/2}}
\exp\left[-\frac12\sum_{lm}\sum_{l'm'}a_{lm}^*(C^{-1})_{lm,l'm'}a_{l'm'}\right]\\  
\nonumber
&\times&
\left\{
1+\frac16\sum_{{\rm all}~l_im_j}
\langle a_{l_1m_1}a_{l_2m_2}a_{l_3m_3}\rangle
\left[
(C^{-1}a)_{l_1m_1}(C^{-1}a)_{l_2m_2}(C^{-1}a)_{l_3m_3}
\right.\right.\\
& &
\left.\left.
-3(C^{-1})_{l_1m_1,l_2m_2}(C^{-1}a)_{l_3m_3}
\right]\right\}.
\label{eq:pdf}
\end{eqnarray}
This formula is useful, as it tells us how to estimate the {\it angular bispectrum},
$\langle a_{l_1m_1}a_{l_2m_2}a_{l_3m_3}\rangle$, optimally from a given
data by maximizing this PDF. In practice,
we usually parametrize the bispectrum using a few parameters (e.g., $f_{\rm
NL}$), and estimate those parameters from the data by maximizing the PDF
with respect to the parameters.

In the limit that the contribution of the connected four-point function
(trispectrum) to the PDF is 
negligible compared to those of the power spectrum and bispectrum,
equation~\ref{eq:pdf} contains all the information on
non-Gaussian fluctuations characterized by the covariance matrix,
$C_{l_1m_1,l_2m_2}=\langle a_{l_1m_1}^*a_{l_2m_2}\rangle$, and the angular
bispectrum, $\langle a_{l_1m_1}a_{l_2m_2}a_{l_3m_3}\rangle$.
This approach can be extended straightforwardly to the trispectrum if
necessary. 
\section{Extracting $f_{\rm NL}$ from the CMB data}
\subsection{General formula}
We have not defined what we mean by ``$f_{\rm NL}$.'' For the moment,
let us loosely define it as the {\it amplitude} of a certain shape of
the angular bispectrum:
\begin{equation}
 \langle a_{l_1m_1}a_{l_2m_2}a_{l_3m_3}\rangle
= 
{\cal G}_{l_1l_2l_3}^{m_1m_2m_3}
\sum_i f_{\rm NL}^{(i)}
b_{l_1l_2l_3}^{(i)},
\label{eq:reducedb}
\end{equation}
where the function, $b_{l_1l_2l_3}^{(i)}$, is called the ``reduced
 angular bispectrum'' \cite{komatsu/spergel:2001} and defines the shape of the
 angular bispectrum for a given model denoted by an index $i$ 
(which may refer to, e.g., ``local,'' ``equilateral,'' ``orthogonal,''
 etc), and  
${\cal G}_{l_1l_2l_3}^{m_1m_2m_3}$ is the so-called Gaunt integral,
defined by 
\begin{equation}
 {\cal G}_{l_1l_2l_3}^{m_1m_2m_3}
\equiv 
\int d^2\hat{\bm{n}}
Y_{l_1m_1}(\hat{\bm{n}})Y_{l_2m_2}(\hat{\bm{n}})Y_{l_3m_3}(\hat{\bm{n}}).
\label{eq:gaunt}
\end{equation}
The physical role of the Gaunt integral is to assure that 
$(l_1,m_1)$, $(l_2,m_2)$, and $(l_3,m_3)$ form a triangle. In the
small-angle limit, the Gaunt integral becomes a 2-d delta function: ${\cal
G}_{l_1l_2l_3}^{m_1m_2m_3}\to
(2\pi)^2\delta^D(\bm{l}_1+\bm{l}_2+\bm{l}_3)$ \cite{hu:2000}.

Given this parametrization, one can then maximize
the PDF given in equation~\ref{eq:pdf} with respect to $f_{\rm
NL}^{(i)}$ by solving $d\ln P/df^{(i)}_{\rm NL}=0$ 
and find the optimal estimator:
\begin{equation}
 f_{\rm NL}^{(i)} = \sum_j (F^{-1})_{ij}S_j.
\label{eq:est}
\end{equation}
One can use this formula to determine multiple amplitudes of angular
bispectra simultaneously.

Here, $S_i$ are given by the data as
\begin{eqnarray}
\nonumber
 S_i &\equiv &
\frac16\sum_{{\rm all}~lm}
{\cal G}_{l_1l_2l_3}^{m_1m_2m_3}b_{l_1l_2l_3}^{(i)}\\
&\times&
\left[
(C^{-1}a)_{l_1m_1}(C^{-1}a)_{l_2m_2}(C^{-1}a)_{l_3m_3}
-3(C^{-1})_{l_1m_1,l_2m_2}(C^{-1}a)_{l_3m_3}
\right],
\end{eqnarray}
where $1/6$ is included such that $F_{ij}$ in equation~\ref{eq:est}
becomes the Fisher matrix of $f_{\rm NL}^{(i)}$. In other words, the
covariance matrix of $f_{\rm NL}^{(i)}$ is given by the inverse of
$F_{ij}$, i.e., 
\begin{equation}
(F^{-1})_{ij}=\langle f_{\rm NL}^{(i)}f_{\rm
NL}^{(j)}\rangle -  \langle f_{\rm NL}^{(i)}\rangle \langle f_{\rm
NL}^{(j)}\rangle. 
\end{equation}
The 68\% uncertainty in $f_{\rm NL}^{(i)}$ is given by 
$\Delta f_{\rm NL}^{(i)}=(F^{-1})_{ii}$.

Using the definition of the Gaunt integral given in
equation~\ref{eq:gaunt}, we rewrite $S_i$ as
\begin{eqnarray}
 S_i &=&
\frac16\int d^2\hat{\bm{n}}
\sum_{l_1l_2l_3}b_{l_1l_2l_3}^{(i)}
\left[
e_{l_1}(\hat{\bm{n}})e_{l_2}(\hat{\bm{n}})e_{l_3}(\hat{\bm{n}})
-
3d_{l_1l_2}(\hat{\bm{n}})e_{l_3}(\hat{\bm{n}})
\right],
\end{eqnarray}
where 
\begin{eqnarray}
\label{eq:emap}
 e_l(\hat{\bm{n}})&\equiv& \sum_{m}(C^{-1}a)_{lm}Y_{lm}(\hat{\bm{n}}),\\
 d_{ll'}(\hat{\bm{n}})&\equiv& \sum_{mm'}(C^{-1})_{lm,l'm'}
Y_{lm}(\hat{\bm{n}})Y_{l'm'}(\hat{\bm{n}}).
\end{eqnarray}
Here, the summation over $m$ can be done using the Fast Fourier Transform
(FFT), as $Y_{lm}(\theta,\phi)\propto e^{im\phi}$. This technique is
used by the {\sf HEALPix} package \cite{gorski/etal:2005}, and thus one
may use {\sf HEALPix} to do this summation.
To compute $d_{ll'}(\hat{\bm{n}})$, one may use Monte Carlo
simulations. Namely, as $C_{lm,l'm'}=\langle a^*_{lm}a_{l'm'}\rangle$, 
we have the exact relation between $d_{ll'}$ and $e_l$:
$d_{ll'}(\hat{\bm{n}}) =\langle
  e_l(\hat{\bm{n}})e_{l'}(\hat{\bm{n}})\rangle$.
One can evaluate the ensemble average using the Monte Carlo
simulation of the CMB and the instrumental noise. Let us denote this
operation by 
$d_{ll'}(\hat{\bm{n}}) =\langle
  e_l(\hat{\bm{n}})e_{l'}(\hat{\bm{n}})\rangle_{\rm MC}$. The final
  formula for $S_i$ is
\begin{eqnarray}
 S_i &=&
\frac16\int d^2\hat{\bm{n}}
\sum_{l_1l_2l_3}b_{l_1l_2l_3}^{(i)}
\left[
e_{l_1}(\hat{\bm{n}})e_{l_2}(\hat{\bm{n}})e_{l_3}(\hat{\bm{n}})
-
3e_{l_3}(\hat{\bm{n}})\langle e_{l_1}(\hat{\bm{n}})e_{l_2}(\hat{\bm{n}})\rangle_{\rm
MC}
\right],
\label{eq:skew}
\end{eqnarray}
which is valid for general forms of $b_{l_1l_2l_3}^{(i)}$. 
Note that the integral, $\int d^2\hat{\bm{n}}$, must be
done over the full sky, even in the presence of the mask: 
the information on the mask is included in the calculation of the Fisher
matrix, $F_{ij}$.
The only
assumptions that we have made so far are: (1) each angular bispectrum
component has only one free parameter, i.e., the amplitude, and (2)
non-Gaussianity (if any) is weak, and the PDF of $a_{lm}$ is given by
equation~\ref{eq:pdf}. 

Finally, the explicit form of the Fisher matrix is given by
\begin{eqnarray}
\nonumber
 F_{ij}&=& \frac{f_{\rm sky}}{6}\sum_{{\rm
  all}~lm}\sum_{{\rm all}~l'm'}{\cal
  G}_{l_1l_2l_3}^{m_1m_2m_3}b_{l_1l_2l_3}^{(i)}\\
& &\times
 (C^{-1})_{l_1m_1,l_1'm_1'}(C^{-1})_{l_2m_2,l_2'm_2'}(C^{-1})_{l_3m_3,l_3'm_3'}b_{l_1'l_2'l_3'}^{(j)}{\cal
 G}_{l_1'l_2'l_3'}^{m_1'm_2'm_3'},
\end{eqnarray}
where $f_{\rm sky}$ is the fraction of the sky outside of the mask.
When the covariance matrix is diagonal, the expression simplifies to 
\begin{equation}
 F_{ij}= \frac{f_{\rm sky}}{6}\sum_{{\rm all}~l}I_{l_1l_2l_3}
\frac{b_{l_1l_2l_3}^{(i)}b_{l_1l_2l_3}^{(j)}}{C_{l_1}C_{l_2}C_{l_3}},
\label{eq:simple}
\end{equation}
where 
\begin{eqnarray}
 I_{l_1l_2l_3}&\equiv& \sum_{{\rm all}~m}({\cal
 G}_{l_1l_2l_3}^{m_1m_2m_3})^2
= 
\frac{(2l_1+1)(2l_2+1)(2l_3+1)}{4\pi}
\left(
\begin{array}{ccc}
l_1&l_2&l_3\\
0&0&0
\end{array}
\right)^2.
\end{eqnarray}
Equation~\ref{eq:simple} may also be written as
\begin{equation}
F_{ij}=f_{\rm sky}\sum_{l_3\le l_2\le l_1} 
I_{l_1l_2l_3}
\frac{b_{l_1l_2l_3}^{(i)}b_{l_1l_2l_3}^{(j)}}{C_{l_1}C_{l_2}C_{l_3}\Delta_{l_1l_2l_3}}, 
 \end{equation}
where $\Delta_{l_1l_2l_3}=1$, 2, and 6 when all of $l_i$'s are
different, two of $l_i'$ are the same, and all of $l_i$'s are the same,
respectively. 

\subsection{Poisson bispectrum}
As a warm up, let us consider the simplest example: point sources
randomly distributed over the sky. As mentioned already, this is a
contamination of the primordial non-Gaussianity parameters, and thus an
accurate measurement of this component is quite important, especially
for {\sl Planck} as well as for the South Pole Telescope (SPT) and Atacama
Cosmology Telescope (ACT), which are working at high frequencies
($\nu>100$~GHz) where
star-forming galaxies dominate $a_{lm}$ at $l>1000$.

For the Poisson distribution, the reduced bispectrum is independent of
multipoles, $b^{\rm src}_{l_1l_2l_3}=1$, in the absence of window
functions, and is given by 
\begin{equation}
 b_{l_1l_2l_3}^{{\rm src}}=w_{l_1}w_{l_2}w_{l_3},
\end{equation}
in the presence of window functions. Here,
$w_l$ is an experimental window function (a product of the beam
transfer function and the pixel window function). Let us then use $b_{\rm
src}$ (instead of $f_{\rm NL}$ because this component has nothing to do
with primordial fluctuations) to denote the amplitude of the Poisson
bispectrum. 

From the data, we measure $S_{\rm src}$ given by
\begin{equation}
 S_{\rm src}
= \frac16\int d^2\hat{\bm{n}}
\left[
E^3(\hat{\bm{n}})
-
3E(\hat{\bm{n}})
\langle E^2(\hat{\bm{n}})\rangle_{\rm MC}
\right],
\end{equation}
where a map $E(\hat{\bm{n}})$ is defined by \cite{komatsu/etal:2009}
\begin{equation}
 E(\hat{\bm{n}})\equiv 
\sum_{l} w_l e_l(\hat{\bm{n}})
=
\sum_{lm} w_l (C^{-1}a)_{lm}Y_{lm}(\hat{\bm{n}}),
\label{eq:E}
\end{equation}
where $e_l(\hat{\bm{n}})$ is given by equation~(\ref{eq:emap}). The
$E(\hat{\bm{n}})$ map is a Wiener-filtered map of point sources randomly
distributed on the sky.

\subsection{Primordial bispectra}

\begin{figure}[t]
\centering \noindent
\includegraphics[width=14cm]{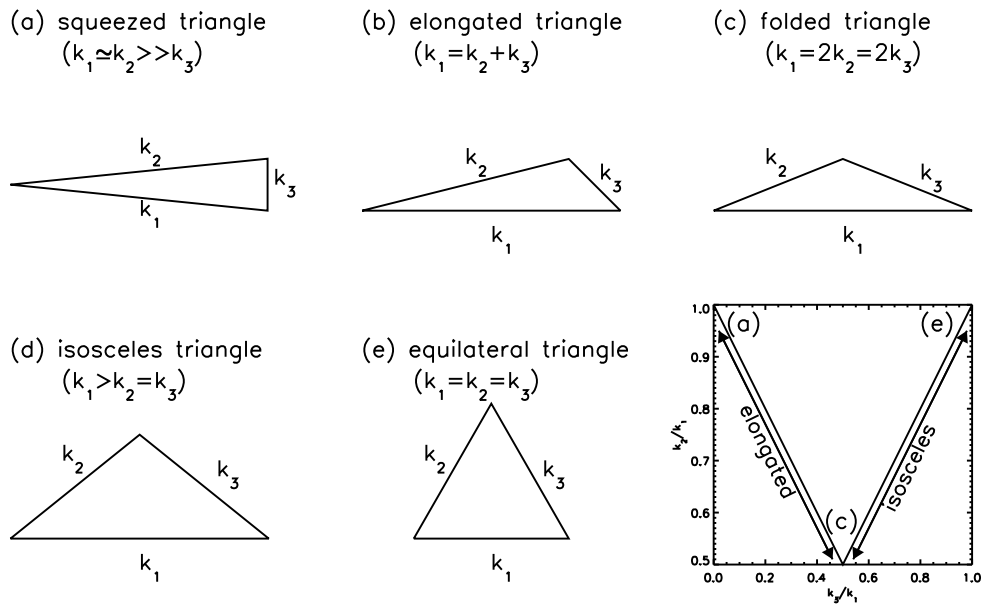}
\caption{%
 Visual representations of triangles forming the bispectrum, 
 $B_\Phi(k_1,k_2,k_3)$, with various combinations of
 wavenumbers satisfying $k_3\le k_2\le k_1$. This figure is adopted from
 \cite{jeong/komatsu:2009}. 
} 
\label{fig:triangles}
\end{figure}

(Most of this subsection is adopted from Section~6.1 of \cite{komatsu/etal:prep}.)
During the period of cosmic inflation
\cite{starobinsky:1979,starobinsky:1982,guth:1981,sato:1981,linde:1982,albrecht/steinhardt:1982},
quantum fluctuations were generated and became the seeds for the cosmic
structures that we observe today
\cite{mukhanov/chibisov:1981,hawking:1982,starobinsky:1982,guth/pi:1982,bardeen/steinhardt/turner:1983}. See
\cite{linde:1990,mukhanov/feldman/brandenberger:1992,liddle/lyth:CIALSS,liddle/lyth:PDP,bassett/tsujikawa/wands:2006,linde:2008}
for reviews.

Inflation predicts that the statistical distribution of primordial
fluctuations is nearly a Gaussian distribution with random phases. 
Measuring deviations from a Gaussian distribution, i.e., non-Gaussian
correlations in primordial fluctuations, is a 
powerful test of inflation, as how precisely the distribution is
(non-)Gaussian depends on the detailed physics of inflation.
See \cite{bartolo/etal:2004,komatsu/etal:astro2010} for reviews.

The observed angular bispectrum is related to the
3-dimensional bispectrum of primordial curvature perturbations,
$\langle\zeta_{\bm{k}_1}\zeta_{\bm{k}_2}\zeta_{\bm{k}_3}\rangle =(2\pi)^3\delta^D(\bm{k}_1+\bm{k}_2+\bm{k}_3)B_\zeta(k_1,k_2,k_3)$. In the linear order, the primordial
curvature perturbation is related to Bardeen's curvature perturbation
\cite{bardeen:1980} in
the matter-dominated era, $\Phi$, by $\zeta=\frac53\Phi$
\cite{kodama/sasaki:1984}. The CMB 
temperature anisotropy in the Sachs--Wolfe limit
\cite{sachs/wolfe:1967} is given by $\Delta
T/T=-\frac13\Phi=-\frac15\zeta$.
We write the bispectrum of $\Phi$ as
\begin{equation}
\langle\Phi({\bm{k}_1})\Phi({\bm{k}_2})\Phi({\bm{k}_3})\rangle =(2\pi)^3\delta^D(\bm{k}_1+\bm{k}_2+\bm{k}_3)F(k_1,k_2,k_3).
\end{equation}

There is a useful way of visualizing the shape
dependence of the bispectrum.
We can study the structure of the bispectrum 
by plotting the magnitude of $F(k_1,k_2,k_3)(k_2/k_1)^2(k_3/k_1)^2$ as a
function of $k_2/k_1$ and $k_3/k_1$ for a given $k_1$, with 
a condition that $k_1\ge k_2\ge k_3$ is satisfied. 
In order to classify various shapes of the triangles, 
let us use the following names:
squeezed ($k_1\simeq k_2\gg k_3$), elongated ($k_1=k_2+k_3$), 
folded ($k_1=2k_2=2k_3$), isosceles ($k_2=k_3$), and equilateral
($k_1=k_2=k_3$). See (a)--(e) of Figure.~\ref{fig:triangles} for the visual
representations of these triangles. 

\begin{figure}[t]
\centering \noindent
\includegraphics[width=15cm]{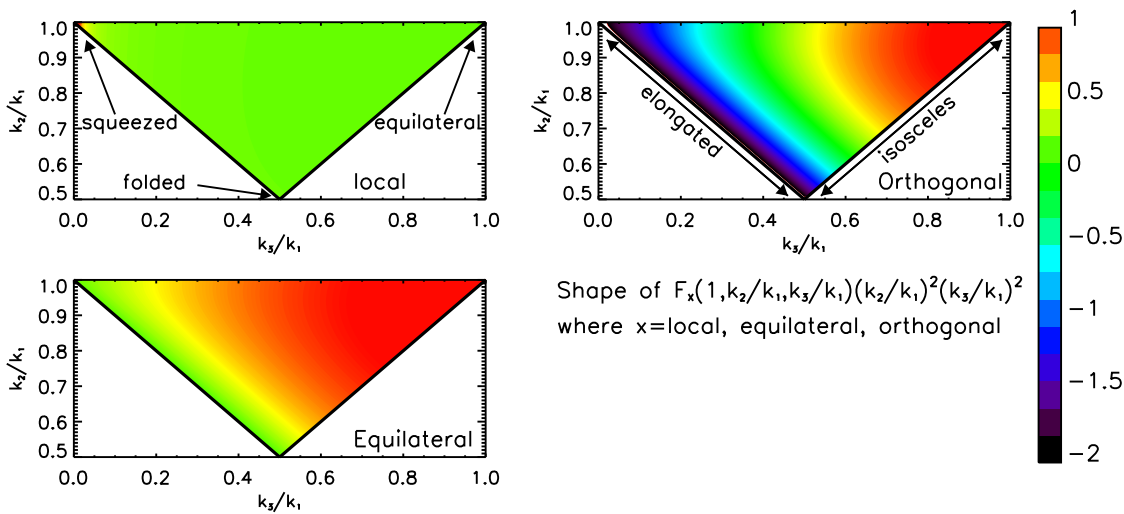}
\caption{%
 Shapes of the primordial bispectra.
 Each panel shows the normalized
 amplitude of $F(k_1,k_2,k_3)(k_2/k_1)^2(k_3/k_1)^2$ as a function of
 $k_2/k_1$ and $k_3/k_1$ 
 for a given $k_1$, with a condition that $k_3\le k_2\le k_1$ is
 satisfied. As the primordial bispectra shown here are (nearly) scale
 invariant, the shapes look similar regardless of the values of $k_1$.
 The amplitude is normalized
 such that it is unity at the point where
 $F(k_1,k_2,k_3)(k_2/k_1)^2(k_3/k_1)^2$ takes on the maximum value. 
(Top Left) The local form given in equation~\ref{eq:Flocal}, which peaks
 at the squeezed configuration.  Note that the most squeezed
 configuration shown here has $k_1=k_2=100k_3$. 
 (Top Right) The orthogonal form given in equation~\ref{eq:Forthog}, which
 has a positive peak at the equilateral configuration, and a negative valley
 along the elongated configurations. 
 (Bottom Left)
 The equilateral form given in equation~\ref{eq:Fequil}, which
 peaks at the equilateral configuration.  Note that all of these shapes are
 nearly orthogonal to each other.
} 
\label{fig:prim}
\end{figure}

We shall explore 3 different shapes of the primordial bispectrum: ``local,''
``equilateral,'' and ``orthogonal.'' They are defined as follows:
\begin{itemize}
 \item [1.] {\bf Local form}. The local form bispectrum is given
       by \cite{gangui/etal:1994,verde/etal:2000,komatsu/spergel:2001}
\begin{eqnarray}
\nonumber
& &F_{\rm local}(k_1,k_2,k_3) \\
\nonumber
&=&
2\fnlKS
	 [P_\Phi(k_1)P_\Phi(k_2)+P_\Phi(k_2)P_\Phi(k_3)
+P_\Phi(k_3)P_\Phi(k_1)]\\
&=& 2A^2\fnlKS\left[\frac1{k^{4-n_s}_1k^{4-n_s}_2}+(2~\mbox{perm}.)\right],
\label{eq:Flocal}
\end{eqnarray}
where $P_\Phi=A/k^{4-n_s}$ is the power spectrum of $\Phi$ with a
       normalization factor $A$. This form is
       called the local form, as this bispectrum can arise from the
       curvature perturbation in the form of
       $\Phi=\Phi_L+\fnlKS\Phi^2_L$, where both sides are evaluated at
       the same location in space ($\Phi_L$ is a linear Gaussian
       fluctuation).\footnote{However, $\Phi=\Phi_L+\fnlKS\Phi^2_L$ is not the
       only way to produce this type of bispectrum. One can also produce
       this form from multi-scalar field inflation models where scalar
       field fluctuations are nearly scale invariant
       \cite{lyth/rodriguez:2005}; multi-scalar models called
       ``curvaton'' scenarios 
       \cite{linde/mukhanov:1997,lyth/ungarelli/wands:2003};
       multi-field models in which one field modulates the decay rate of
       inflaton field
       \cite{dvali/gruzinov/zaldarriaga:2004b,dvali/gruzinov/zaldarriaga:2004a,zaldarriaga:2004};
       multi-field models in which a violent production of particles and
       non-linear reheating,  
       called ``preheating,'' occur due to parametric resonances
       \cite{enqvist/etal:2005,jokinen/mazumdar:2006,chambers/rajantie:2007,bond/etal:2009};
       models in which the universe contracts first and then bounces
       \cite{lehners:2008}.} The local form, $F_{\rm
       local}(k_1,k_2,k_3)(k_2/k_1)^2(k_3/k_1)^2$, peaks at the so-called
       ``squeezed'' triangle for which $k_3\ll k_2\approx k_1$
       \cite{babich/creminelli/zaldarriaga:2004}. See the top-left panel
       of Figure~\ref{fig:prim}. In this limit, we
       obtain \begin{equation}\label{eq:1}
 F_{\rm local}(k_1,k_1,k_3\to 0)=4\fnlKS P_\Phi(k_1)P_\Phi(k_3).
\end{equation}
How large is $\fnlKS$ from inflation?  
The earlier calculations showed that $\fnlKS$ from single-field slow-roll
inflation would be of order the slow-roll parameter, $\epsilon\sim
10^{-2}$
\cite{salopek/bond:1990,falk/rangarajan/srendnicki:1993,gangui/etal:1994}. 
More recently, Maldacena \cite{maldacena:2003} and Acquaviva et
       al. \cite{acquaviva/etal:2003} 
       found that the coefficient of $P_\Phi(k_1)P_\Phi(k_3)$ from the simplest
single-field slow-roll inflation with the canonical kinetic term
in the squeezed limit is given by 
\begin{equation}
 F_{\rm local}(k_1,k_1,k_3\to 0)=\frac53(1-n_s) P_\Phi(k_1)P_\Phi(k_3).
\label{eq:singleprediction}
\end{equation}
Comparing this result with the form predicted by the $\fnlKS$ model,
one obtains $\fnlKS=(5/12)(1-n_s)$, which gives $\fnlKS=0.015$ for
       $n_s=0.963$. 
 \item [2.] {\bf Equilateral form}. The equilateral form bispectrum is given
       by \cite{creminelli/etal:2006}
\begin{eqnarray}
\nonumber
& &F_{\rm equil}(k_1,k_2,k_3)\\
\nonumber
&=& 6A^2\fnleq
\left\{
-\frac1{k^{4-n_s}_1k^{4-n_s}_2}-\frac1{k^{4-n_s}_2k^{4-n_s}_3}
-\frac1{k^{4-n_s}_3k^{4-n_s}_1}\right.\\
& &\left.
-\frac2{(k_1k_2k_3)^{2(4-n_s)/3}}
+\left[\frac1{k^{(4-n_s)/3}_1k^{2(4-n_s)/3}_2k^{4-n_s}_3}
+\mbox{(5 perm.)}\right]\right\}.
\label{eq:Fequil}
\end{eqnarray}
This function approximates the bispectrum forms that arise from a class of
       inflation models in which scalar fields have 
       non-canonical kinetic terms. One example is the so-called
       Dirac-Born-Infeld (DBI) inflation
       \cite{silverstein/tong:2004,alishahiha/silverstein/tong:2004},
       which gives $\fnleq\propto -1/c_s^2$ in the limit of $c_s\ll
       1$, where $c_s$ is the effective sound speed at which scalar
       field fluctuations propagate relative to the speed of
       light. There are various other models that 
       can produce $\fnleq$ 
       \cite{arkani-hamed/etal:2004,seery/lidsey:2005,chen/etal:2007,cheung/etal:2008,li/wang/wang:2008}. The equilateral form, $F_{\rm
       equil}(k_1,k_2,k_3)(k_2/k_1)^2(k_3/k_1)^2$, peaks at the
       equilateral configuration for which $k_1=k_2=k_3$. See the
       bottom-left panel 
       of Figure~\ref{fig:prim}.
       The local and equilateral forms are nearly orthogonal to each
       other, which means that both can be measured nearly
       independently. 
 \item [3.] {\bf Orthogonal form}. The orthogonal form, which is
       constructed such that it is nearly orthogonal to both the local
       and equilateral forms, is given by
       \cite{senatore/smith/zaldarriaga:2010}
\begin{eqnarray}
\nonumber
& &F_{\rm orthog}(k_1,k_2,k_3)\\
\nonumber
&=& 6A^2\fnlor
\left\{
-\frac3{k^{4-n_s}_1k^{4-n_s}_2}-\frac3{k^{4-n_s}_2k^{4-n_s}_3}
-\frac3{k^{4-n_s}_3k^{4-n_s}_1}\right.\\
& &
\left.
-\frac8{(k_1k_2k_3)^{2(4-n_s)/3}}
+\left[\frac3{k^{(4-n_s)/3}_1k^{2(4-n_s)/3}_2k^{4-n_s}_3}
 +\mbox{(5 perm.)}\right]\right\}.
\label{eq:Forthog}
\end{eqnarray}
This form approximates the forms that arise from a linear combination of
       higher-derivative scalar-field interaction terms, each of which
       yields forms similar to the equilateral shape. 
Senatore, Smith and Zaldarriaga \cite{senatore/smith/zaldarriaga:2010}
       found that, using the ``effective 
       field theory of inflation'' approach \cite{cheung/etal:2008},
       a certain linear combination of similarly equilateral shapes can
       yield a distinct shape which is orthogonal to both the local and
       equilateral forms. The orthogonal form, $F_{\rm
       orthog}(k_1,k_2,k_3)(k_2/k_1)^2(k_3/k_1)^2$, 
       has a positive peak at the equilateral configuration, and a
       negative valley along the elongated configurations. See the
       top-right panel 
       of Figure~\ref{fig:prim}.
\end{itemize}
Note that these are not the most general forms one can write down, and
there are other forms which would probe different aspects of the physics
of inflation
\cite{moss/chun:2007,moss/graham:2007,chen/etal:2007,holman/tolley:2008,chen/wang:prep,chen/wang:prepb}.  

Of these forms, the local form bispectrum has special significance. 
Creminelli and Zaldarriaga \cite{creminelli/zaldarriaga:2004} showed that 
not only models with the canonical kinetic term, but
{\it all}
single-inflation models predict the
bispectrum in the squeezed limit given by
equation~\ref{eq:singleprediction}, regardless of the form of potential,
kinetic term,
slow-roll, or initial vacuum state.
Also see \cite{seery/lidsey:2005,chen/etal:2007,cheung/etal:2008}.
This means that a convincing detection of $\fnlKS$ would rule out {\it
all} single-field inflation models. 

\subsection{Optimal estimator for $\fnlKS$}
Given the form of $\Phi$, one can calculate the harmonic coefficients of
temperature and $E$-mode polarization anisotropies as
\begin{eqnarray}
\label{eq:gT}
 a^T_{lm} &=& 4\pi(-i)^l\int \frac{d^3\bm{k}}{(2\pi)^3}
\Phi(\bm{k})g_{Tl}(k)Y_{lm}^*(\bm{k}),\\
\label{eq:gP}
 a^E_{lm} &=&  4\pi(-i)^l\sqrt{\frac{(l+2)!}{(l-2)!}}\int
  \frac{d^3\bm{k}}{(2\pi)^3} 
\Phi(\bm{k})g_{Pl}(k)Y_{lm}^*(\bm{k}),
\end{eqnarray}
where $g_{Tl}(k)$ and $g_{Pl}(k)$ are the radiation transfer functions
of the temperature and polarization anisotropies, respectively, which can be
calculated by solving the linearized Boltzmann equations.
One may use the publicly-available Boltzmann codes such as {\sf CMBFAST}
\cite{seljak/zaldarriaga:1996} 
or {\sf CAMB} \cite{lewis/challinor/lasenby:2000} for computing the
radiation transfer functions.\footnote{A {\sf CMBFAST}-based code for
computing $g_{Tl}(k)$ and $g_{Pl}(k)$ is available at 
\href{http://gyudon.as.utexas.edu/~komatsu/CRL}{http://gyudon.as.utexas.edu/$\sim$komatsu/CRL}.
A recent version of {\sf
CAMB} has an option to calculate these functions (\href{http://camb.info}{http://camb.info}).}

From now on, we shall focus on the temperature anisotropy, largely for
simplicity.
(See \cite{babich/zaldarriaga:2004,yadav/komatsu/wandelt:2007} for the
treatment of polarization in the angular bispectrum.)
The limits on $f_{\rm NL}$ expected from {\sl Planck} are dominated by
the temperature information, and thus the polarization information is
not expected to yield competitive limits over the next, say, $>5$ years.

For the local-form bispectrum given in equation~\ref{eq:Flocal}, the
reduced bispectrum (equation~\ref{eq:reducedb}) is given by
\cite{komatsu/spergel:2001}
\begin{equation}
 b_{l_1l_2l_3}^{\rm local}
=
2\int r^2dr\left[
\beta_{l_1}(r)\beta_{l_2}(r)\alpha_{l_3}(r)
+(2~{\rm perm}.)
\right]w_{l_1}w_{l_2}w_{l_3},
\end{equation}
where
\begin{eqnarray}
 \alpha_l(r)&=&\frac2{\pi}\int k^2dk~g_{Tl}(k)j_l(kr),\\
 \beta_l(r)&=&\frac2{\pi}\int k^2dk~P_\Phi(k)g_{Tl}(k)j_l(kr).
\end{eqnarray}
Using this form in equation~\ref{eq:skew}, one finds $S_{\rm local}$ as
\begin{eqnarray}
\nonumber
 S_{\rm local}
&=&
\int r^2dr
\int d^2\hat{\bm{n}}
\left[
A(\hat{\bm{n}},r)B^2(\hat{\bm{n}},r)
-2B(\hat{\bm{n}})\langle A(\hat{\bm{n}},r)B(\hat{\bm{n}},r) \rangle_{\rm
MC}\right.\\
&  &\left.
-A(\hat{\bm{n}},r)\langle B^2(\hat{\bm{n}},r) \rangle_{\rm
MC}
\right],
\end{eqnarray}
which can be measured from the data. Here, maps $A(\hat{\bm{n}},r)$ and 
$B(\hat{\bm{n}},r)$ are defined by \cite{komatsu/spergel/wandelt:2005}
\begin{eqnarray}
 A(\hat{\bm{n}},r) &\equiv &
\sum_{l} w_l\alpha_l(r) e_l(\hat{\bm{n}})
=
\sum_{lm} w_l\alpha_l(r) (C^{-1}a)_{lm}Y_{lm}(\hat{\bm{n}}),\\
B(\hat{\bm{n}},r) &\equiv &
\sum_{l} w_l\beta_l(r) e_l(\hat{\bm{n}})
=
\sum_{lm} w_l\beta_l(r) (C^{-1}a)_{lm}Y_{lm}(\hat{\bm{n}}),
\end{eqnarray}
where 
$e_l(\hat{\bm{n}})$ is given by equation~(\ref{eq:emap}).

\subsection{Optimal estimator for $\fnleq$}
For the equilateral-form bispectrum given in equation~\ref{eq:Fequil}, the
reduced bispectrum is given by
\cite{creminelli/etal:2006}
\begin{eqnarray}
\nonumber
 b_{l_1l_2l_3}^{\rm equil}
&=&
-3b_{l_1l_2l_3}^{\rm local}
+6\int r^2dr\left[
\beta_{l_1}(r)\gamma_{l_2}(r)\delta_{l_3}(r)+(5~{\rm perm}.)\right.\\
& &\left.-2\delta_{l_1}(r)\delta_{l_2}(r)\delta_{l_3}(r)
\right]w_{l_1}w_{l_2}w_{l_3},
\end{eqnarray}
where
\begin{eqnarray}
 \gamma_l(r)&=&\frac2{\pi}\int k^2dk~P_\Phi^{1/3}(k)g_{Tl}(k)j_l(kr),\\
 \delta_l(r)&=&\frac2{\pi}\int k^2dk~P_\Phi^{2/3}(k)g_{Tl}(k)j_l(kr).
\end{eqnarray}
Using this form in equation~\ref{eq:skew}, one finds $S_{\rm equil}$ as
\begin{eqnarray}
\nonumber
 S_{\rm equil}
&=&
-3S_{\rm local}
+6
\int r^2dr
\int d^2\hat{\bm{n}}
\left\{
B(\hat{\bm{n}},r)C(\hat{\bm{n}},r)D(\hat{\bm{n}},r)\right.\\
\nonumber
& &
-B(\hat{\bm{n}})\langle C(\hat{\bm{n}},r)D(\hat{\bm{n}},r) \rangle_{\rm
MC}
-C(\hat{\bm{n}})\langle B(\hat{\bm{n}},r)D(\hat{\bm{n}},r) \rangle_{\rm
MC}\\
\nonumber
&  &\left.
-D(\hat{\bm{n}})\langle B(\hat{\bm{n}},r)C(\hat{\bm{n}},r) \rangle_{\rm
MC}
-\frac13\left[D^3(\hat{\bm{n}},r)-3D(\hat{\bm{n}},r)\langle
	 D^2(\hat{\bm{n}},r)\rangle_{\rm MC}
\right]\right\},\\
\end{eqnarray}
which can be measured from the data. Here, maps $C(\hat{\bm{n}},r)$ and 
$D(\hat{\bm{n}},r)$ are defined by \cite{creminelli/etal:2006}
\begin{eqnarray}
 C(\hat{\bm{n}},r) &\equiv &
\sum_{l} w_l\gamma_l(r) e_l(\hat{\bm{n}})
=
\sum_{lm} w_l\gamma_l(r) (C^{-1}a)_{lm}Y_{lm}(\hat{\bm{n}}),\\
D(\hat{\bm{n}},r) &\equiv &
\sum_{l} w_l\delta_l(r) e_l(\hat{\bm{n}})
=
\sum_{lm} w_l\delta_l(r) (C^{-1}a)_{lm}Y_{lm}(\hat{\bm{n}}).
\end{eqnarray}

\subsection{Optimal estimator for $\fnlor$}
For the equilateral-form bispectrum given in equation~\ref{eq:Forthog}, the
reduced bispectrum is given by
\cite{senatore/smith/zaldarriaga:2010}
\begin{eqnarray}
 b_{l_1l_2l_3}^{\rm orthog}
&=&
3b_{l_1l_2l_3}^{\rm equil}
-12\int
r^2dr~\delta_{l_1}(r)\delta_{l_2}(r)\delta_{l_3}(r)w_{l_1}w_{l_2}w_{l_3}. 
\end{eqnarray}
Using this form in equation~\ref{eq:skew}, one finds $S_{\rm orthog}$ as
\begin{eqnarray}
 S_{\rm orthog}
&=&
3S_{\rm equil}
-2
\int r^2dr
\int d^2\hat{\bm{n}}
\left[D^3(\hat{\bm{n}},r)-3D(\hat{\bm{n}},r)\langle
	 D^2(\hat{\bm{n}},r)\rangle_{\rm MC}
\right],
\end{eqnarray}
which can be measured from the data.

\section{Secondary anisotropy}
\subsection{General formula for the lensing-secondary coupling}

Given the special importance of the local-form bispectrum, we must
understand what other (non-primordial) effects might also produce the
local form, potentially preventing us from measuring $\fnlKS$. 

The local-form bispectrum is generated when the power spectrum of
short-wavelength fluctuations is modulated by long-wavelength
fluctuations; thus, a mechanism that couples small scales to large
scales can potentially generate the local-form bispectrum.

The weak gravitational lensing provides one such mechanism. The
local-form bispectrum may then be generated when long- and short-wavelength
fluctuations are coupled by the lensing. To see how this might happen,
let us write the observed temperature anisotropy in terms of the
original (unlensed) contribution from the last scattering surface at $z=1090$,
$\Delta T^P$ (where ``$P$'' stands for ``{\it primary}''), the lensing potential, $\phi$, and the secondary
anisotropy generated between $z=1090$ and $z=0$, $\Delta T^S$ (where
``$S$'' stands for ``{\it secondary}''):
\begin{eqnarray}
\nonumber
 \Delta T(\hat{\bm{n}})
&=&\Delta T^P(\hat{\bm{n}}+\vec{\partial}\phi)+\Delta
T^S(\hat{\bm{n}})\\
&\approx&
\Delta T^P(\hat{\bm{n}})
+[(\vec{\partial}\phi)\cdot(\vec{\partial}\Delta T^P)](\hat{\bm{n}})
+\Delta T^S(\hat{\bm{n}}),
\end{eqnarray}
where
\begin{equation}
 \phi(\hat{\bm{n}})=-2\int_0^{r_*}dr
\frac{r_*-r}{rr_*}\Phi(r,\hat{\bm{n}}r),
\end{equation}
and $r_*$ is the comoving distance out to $z=1090$, and 
$\Phi$ is Bardeen's curvature perturbation, which is related to the
usual Newtonian gravitational potential by $\Phi=-\Phi_{\rm Newton}$.
Transforming this into harmonic space and computing the reduced
bispectrum, one obtains \cite{goldberg/spergel:1999}
\begin{equation}
 b_{l_1l_2l_3}^{\rm{lens-}S}
=
\left[
\frac{l_1(l_1+1)-l_2(l_2+1)+l_3(l_3+1)}2C_{l_1}^P
C_{l_3}^{\phi S}
+(5~{\rm perm}.)\right]w_{l_1}w_{l_2}w_{l_3},
\label{eq:lens}
\end{equation}
where $C_l^P$ is the power spectrum of the CMB {\it from the decoupling
epoch only} (i.e., no ISW),
and $C_l^{\phi S}\equiv
\langle\phi_{lm}^*a_{lm}^S\rangle$ is the lensing-secondary
cross-correlation power spectrum.

From this result, one finds that a non-zero bispectrum is generated when
the secondary anisotropy traces the large-scale structure (i.e.,
$\Phi$).
Various secondary effects have been studied in the literature:
the Sunyaev--Zel'dovich effect \cite{goldberg/spergel:1999}, 
cosmic reionization \cite{cooray/hu:2000}, point sources
\cite{babich/pierpaoli:2008}, and ISW \cite{goldberg/spergel:1999}.
It has been shown that the last one, the ISW-lensing coupling, is the
most dominant contamination of $\fnlKS$
\cite{serra/cooray:2008}.

Using equation~\ref{eq:lens} in equation~\ref{eq:skew}, 
 one finds $S_{{\rm lens-}S}$ as
\begin{eqnarray}
\nonumber
 S_{{\rm lens-}S}
&=& \frac12
\int d^2\hat{\bm{n}}
\left\{P(\hat{\bm{n}})[\partial^2 E](\hat{\bm{n}})Q(\hat{\bm{n}})\right.\\
& &-[\partial^2 P](\hat{\bm{n}})E(\hat{\bm{n}})Q(\hat{\bm{n}})
-P(\hat{\bm{n}})E(\hat{\bm{n}})[\partial^2 Q](\hat{\bm{n}})\\
& &\left.
+(\mbox{linear terms})
\right\},
\label{eq:lens-S}
\end{eqnarray}
which can be measured from the data. 
Here, the ``linear terms'' contain 9 terms with $\langle\rangle_{\rm
MC}$, such as $-P(\hat{\bm{n}})\langle[\partial^2
E](\hat{\bm{n}})Q(\hat{\bm{n}})\rangle_{\rm MC}$, etc.
The map $E(\hat{\bm{n}})$ is given by equation~\ref{eq:E},
and the other maps are defined by
\begin{eqnarray}
 P(\hat{\bm{n}})&\equiv&
\sum_l w_l C_l^P e_l(\hat{\bm{n}})
=
\sum_{lm} w_l C_l^P (C^{-1}a)_{lm}Y_{lm}(\hat{\bm{n}}),\\
Q(\hat{\bm{n}})&\equiv&
\sum_l w_l C_l^{\phi S} e_l(\hat{\bm{n}})
=
\sum_{lm} w_l C_l^{\phi S} (C^{-1}a)_{lm}Y_{lm}(\hat{\bm{n}}).
\end{eqnarray}
The maps with $\partial^2$ are given by
$\partial^2 P = -\sum_{l}l(l+1)w_l C_l^P e_l(\hat{\bm{n}})$, etc.
The map $P(\hat{\bm{n}})$ is a Wiener-filtered map of the primary
temperature anisotropy from $z=1090$. 

\subsection{Lensing-ISW coupling}
A change in the curvature perturbation yields a secondary temperature
anisotropy via the ISW effect \cite{sachs/wolfe:1967}:
\begin{equation}
 \frac{\Delta
  T^{\rm ISW}(\hat{\bm{n}})}{T}=-2\int_0^{r_*}dr~\frac{\partial
  \Phi}{\partial r}(r,\hat{\bm{n}}r),
\end{equation}
where $r$ is the comoving distance and $r_*$ is the comoving distance
out to $z=1090$. Here, note again $\Phi=-\Phi_{\rm Newton}$.
The cross-power spectrum of $\phi$ and the ISW effect is then given by
\begin{equation}
    \label{eq:clphi}
    C_l^{\phi,{\rm ISW}}
    = 
    4\int_{0}^{r_*} dr\frac{r_*-r}{r_* r^3} P_{\Phi\Phi'}\left(\frac{l}{r},r\right),
\end{equation}
where $P_{\Phi\Phi'}(k,r)$ is the cross-power spectrum of $\Phi$ and
$\Phi'\equiv \partial\Phi/\partial r$, which can be calculated from the
power spectrum of $\Phi$, $P_\Phi(k,r)$, as
$P_{\Phi\Phi'}(k,r)=\frac12[\partial P_{\Phi}(k,r)/\partial r]$
\cite{verde/spergel:2002,nishizawa/etal:2008}.
Here, $P_\Phi(k,r)$ is not the primordial power spectrum, but it
includes the linear transfer function, $T(k)$, and the growth factor of
$\Phi$, $g(r)$: 
\begin{equation}
 P_\Phi(k,r) = \frac{A}{k^{4-n_s}}\left[T(k)g(r)\right]^2.
\end{equation}
Using this, one finds $P_{\Phi\Phi'}(k,r)=({g'}/{g})P_\Phi(k,r)$.
Note that $g(r)$ is normalized such that $g(r)=1$ during the
matter-dominated era. 

With this result, it is easy to see why the lensing-ISW coupling yields
the squeezed configuration: on very large scales, where $T(k)\to 1$,
$C_l^{\phi,{\rm ISW}}\propto 1/l^3$. On smaller scales, $T(k)$ declines
with $k$, and thus $C_l^{\phi,{\rm ISW}}$ falls faster than $1/l^3$. The
lensing coupling includes $l(l+1)C_l^{\phi,{\rm ISW}}$, which falls
faster than $1/l$, i.e., the largest power comes from the smallest $l$. 

A recent estimate by Hanson et al. \cite{hanson/etal:2009} showed 
that the lensing-ISW coupling, {\it if not included in the parameter
estimation}, would 
bias $\fnlKS$ by $\Delta \fnlKS=9.3$. 
The expected bias for {\sl WMAP} is $\Delta \fnlKS=2.7$
\cite{komatsu/etal:prep}. 
One can remove this bias by including the lensing-ISW coupling (or any
other lensing-secondary couplings) using the optimal estimator given by equation~\ref{eq:lens-S}.

\section{Second-order effect}
\subsection{General discussion}
So far, we have assumed that one can use equation~\ref{eq:gT}:
$$
 a_{lm} = 4\pi(-i)^l\int \frac{d^3\bm{k}}{(2\pi)^3}
\Phi_p(\bm{k})g_{Tl}(k)Y_{lm}^*(\bm{k}),
$$
to convert the primordial curvature perturbation to the temperature
anisotropy. (Here, the subscript ``p'' stands for ``{\it primordial},''
by which we mean $\Phi_p=\frac35\zeta$ without the linear transfer function.)
However, this equation is valid only for linear theory.
As {\it any} non-linear effects can produce non-Gaussianity, one has to
study the impacts of various non-linear effects on the observed
non-Gaussianity. 

The origin of the linear radiation transfer function is the linearized
Boltzmann equation:
\begin{equation}
 \frac{\partial{\Delta^{(1)}}}{\partial\eta}
+ ik\mu\Delta^{(1)} 
+\sigma_Tn_ea\Delta^{(1)}
=  S^{(1)}(k,\mu,\eta),
\end{equation}
where $\eta$ is the conformal time, $\mu\equiv
{\hat{\bm{k}}}\cdot{\hat{\bm{n}}}$, $\Delta^{(1)}\equiv 4[\Delta
T^{(1)}(k,\mu,\eta)/T]$ is the perturbation in the photon energy
density, and $S^{(1)}$ is the {\it linear source function}, 
which depends on the metric perturbations as well as on the density,
velocity, pressure, and stress perturbations of matter and radiation in
the universe and the photon polarization.

The second-order Boltzmann equation is then similarly written as
\begin{equation}
 \frac{\partial{\Delta^{(2)}}}{\partial\eta}
+ ik\mu\Delta^{(2)} 
+\sigma_Tn_ea\Delta^{(2)}
=  S^{(2)}(\bm{k},\hat{\bm{n}},\eta),
\end{equation}
where $\Delta^{(2)}\equiv 8[\Delta
T^{(2)}(\bm{k},\hat{\bm{n}},\eta)/T]+12[\Delta
T^{(1)}(k,\mu,\eta)/T]^2$, and 
$S^{(2)}$ is the {\it second-order source function}. Note that the
azimuthal symmetry is lost at the second order, and thus the
perturbations depend on the directions of $\bm{k}$ and $\hat{\bm{n}}$
independently. In this case, the second-order $a_{lm}$ is given by
\cite{nitta/etal:2009}
\begin{eqnarray}
\nonumber
 a_{lm}^{(2)} &=&
\frac{4\pi}{8}(-i)^l
\int \frac{d^3\bm{k}}{(2\pi)^3}
\int \frac{d^3\bm{k}'}{(2\pi)^3}
\int {d^3\bm{k}''}\delta^D(\bm{k}'+\bm{k}''-\bm{k})
\Phi^{(1)}_p(\bm{k}')\Phi^{(1)}_p(\bm{k}'')\\
& &\times \sum_{l'm'}F_{lm}^{l'm'}(\bm{k}',\bm{k}'',\bm{k})
Y_{l'm'}^*(\hat{\bm{k}}),
\label{eq:nitta}
\end{eqnarray} 
where $F_{lm}^{l'm'}$ is the second-order radiation transfer function,
whose form is determined by the second-order source function, $S^{(2)}$,
in the Boltzmann equation.

The shape of the second-order bispectrum, $\langle
a_{l_1m_2}^{(1)}a_{l_2m_3}^{(1)}a_{l_3m_3}^{(2)}\rangle$, is determined
by the shape of the second-order radiation transfer function.
If the second-order radiation transfer function vanishes in the squeezed
limit, i.e., $F_{lm}^{l'm'}(\bm{k}',\bm{k}'',\bm{k})\to 
0$ for $\bm{k}\to 0$, then the  CMB bispectrum would not peak
at the squeezed configuration, and thus the resulting $\fnlKS$ would be small.

The second-order source function is quite complicated
\cite{bartolo/matarrese/riotto:2006,bartolo/matarrese/riotto:2007,khatri/wandelt:2009,khatri/wandelt:prep,senatore/tassev/zaldarriaga:2009,senatore/tassev/zaldarriaga:2009b,pitrou:2009,pitrou:2009b,pitrou/uzan/bernardeau:2008,pitrou/uzan/bernardeau:prep,beneke/fidler:prep},
but it can be divided into two parts\footnote{This 
decomposition is not gauge invariant, and thus which terms belong to (1)
or (2) depends on the gauge that one chooses. Therefore, one must
specify the gauge when making such a decomposition. Our discussion in
this section is based on the gauge choice made by Bartolo, Matarrese and Riotto
\cite{bartolo/matarrese/riotto:2006,bartolo/matarrese/riotto:2007} and
Pitrou, Uzan and Bernardeau
\cite{pitrou/uzan/bernardeau:2008,pitrou/uzan/bernardeau:prep}, 
which reduces to the Newtonian gauge at the linear order. This seems a
convenient gauge, as the products of the first-order terms only give
$|\fnlKS|<1$ \cite{nitta/etal:2009}.}: 
\begin{itemize}
 \item[(1)] The terms given by the products of the first-order perturbations,
such as $[\Phi^{(1)}]^2$.
 \item[(2)] The terms given by the ``intrinsically second-order terms,''
such as $\Phi^{(2)}$. 
\end{itemize}
The intrinsically second-order terms are sourced
by products of the first-order perturbations, and thus it is created by
the late-time evolution of cosmological perturbations, whereas the terms
in (1) are set by the initial conditions.

The contamination of $\fnlKS$ due to the terms in (1) is small, $|\fnlKS|<1$
\cite{nitta/etal:2009}. Recently, Pitrou, Bernardeau and Uzan
\cite{pitrou/uzan/bernardeau:prep} reported a surprising result that the terms in
(2) would give $\fnlKS\sim 5$ for the {\sl Planck} data ($l_{\rm max}=2000$).

Why surprising?  As the intrinsically second-order terms arise as a
consequence of the late-time evolution of the cosmological
perturbations, they are generated by the causal mechanism, i.e., gravity
and hydrodynamics. It is difficult for the causal mechanism to generate
the bispectrum in the squeezed configuration, as it requires very long
wavelength perturbations to be coupled to short wavelength ones. 

\subsection{Newtonian calculation}
As an example, let us consider the well-known second-order solution for
$\Phi^{(2)}$ 
in the {\it sub-horizon limit}, i.e.,
$k\gg aH$, which is equivalent to taking the non-relativistic (Newtonian)
limit. Here, the second-order Bardeen curvature perturbation is defined
by $\Phi=\Phi^{(1)}+\frac12\Phi^{(2)}$.
The explicit solution is
\cite{pitrou/uzan/bernardeau:2008}\footnote{Note
that our $\Phi$ is $(-1)$ times $\Phi$ used in equation~72 of
\cite{pitrou/uzan/bernardeau:2008}.} 
\begin{eqnarray}
\nonumber
 \frac12\Phi^{(2)}(\bm{k},\eta)&=& 
\frac16\int\frac{d^3\bm{k}'}{(2\pi)^3}d^3\bm{k}''
\delta^D(\bm{k}'+\bm{k}''-\bm{k})
\left(\frac{k'k''\eta}{k}\right)^2\\
& &\times
F_2^{(s)}(\bm{k}',\bm{k}'')
\Phi^{(1)}(\bm{k}')\Phi^{(1)}(\bm{k}''),
\label{eq:newton}
\end{eqnarray}
where the linear perturbation, $\Phi^{(1)}$, on the right hand side is
constant during the matter dominated era, and the symmetrized function,
$F_2^{(s)}$, is defined as
\begin{equation}
F_2^{(s)}(\bm{k}_1,\bm{k}_2)
=\frac{5}{7}
+\frac{\bm{k}_1\cdot\bm{k}_2}{2k_1k_2}
\left(
\frac{k_1}{k_2}+\frac{k_2}{k_1}
\right)
+\frac{2}{7}\left(\frac{\bm{k}_1\cdot\bm{k}_2}{k_1k_2}\right)^2.
\label{eq:F2}
\end{equation}
Note that $F_2^{(s)}$ is related to the function $G$ given in
equation~(8.9) of \cite{bartolo/matarrese/riotto:2007} as 
$G(\bm{k}_1,\bm{k}_2,\bm{k})=-\frac{14}3\left(\frac{k_1k_2}{k}\right)^2
F_2^{(s)}(\bm{k}_1,\bm{k}_2)$ with $\bm{k}=\bm{k}_1+\bm{k}_2$.

\begin{figure}[t]
\centering \noindent
\includegraphics[width=15cm]{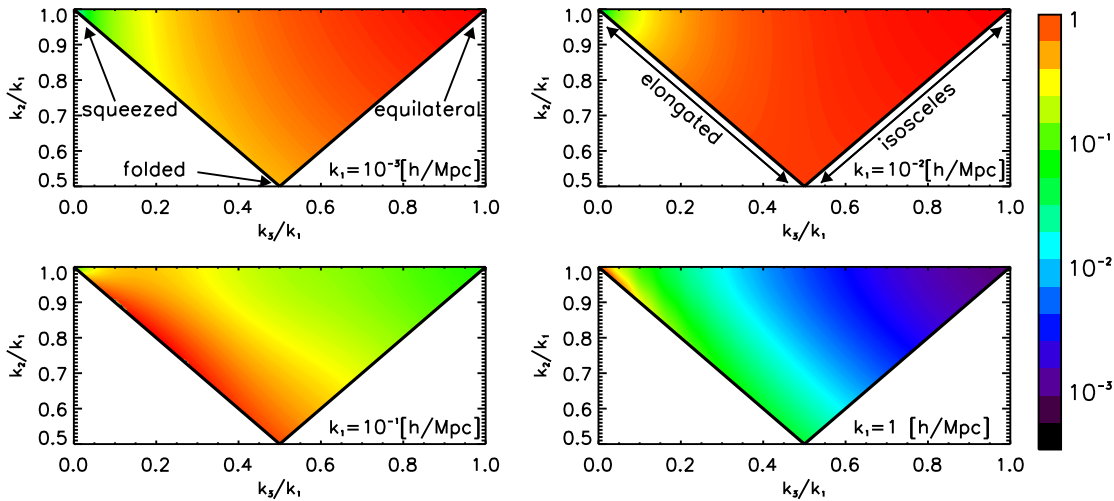}
\caption{%
 Shapes of the second-order bispectrum due to the second-order curvature
 perturbations in the Newtonian limit given in equation~\ref{eq:2nd}.
 Each panel shows the normalized
 amplitude of $F_{\rm 2nd}(k_1,k_2,k_3)(k_2/k_1)^2(k_3/k_1)^2$ as a function of
 $k_2/k_1$ and $k_3/k_1$ 
 for a given $k_1$, with a condition that $k_3\le k_2\le k_1$ is
 satisfied. The amplitude is normalized
 such that it is unity at the point where
 $F_{\rm 2nd}(k_1,k_2,k_3)(k_2/k_1)^2(k_3/k_1)^2$ takes on the maximum value. 
 (Top Left) $k_1=10^{-3}~h~{\rm Mpc}^{-1}$. (Top Right)
 $k_1=10^{-2}~h~{\rm Mpc}^{-1}$. 
 (Bottom Left) $k_1=10^{-1}~h~{\rm Mpc}^{-1}$. (Bottom Right)
 $k_1=1~h~{\rm Mpc}^{-1}$. The CMB data are sensitive to $k_1<k_{\rm
 max}\sim 0.2~h~{\rm Mpc}^{-1}(l_{\rm max}/2000)$, where the
 second-order bispectrum peaks at the equilateral configuration on large
 scales, and peaks along the elongated configurations on a smaller scale 
 ($k_1\sim 0.1~h~{\rm Mpc}^{-1}$).
 On a very small scale ($k_1\sim 1~h~{\rm
 Mpc}^{-1}$), it peaks at the squeezed configuration.
 Note that the most squeezed configuration shown here has $k_1=k_2=100k_3$.
} 
\label{fig:2nd}
\end{figure}

The function $F_2^{(s)}(\bm{k}_1,\bm{k}_2)$ vanishes in the squeezed limit,
$\bm{k}_1=-\bm{k}_2$, and thus the CMB bispectrum generated from
$\Phi^{(2)}$ {\it in the Newtonian limit} is not given by the local
form. To see this, let us calculate
\begin{equation}
 \langle\Phi(\bm{k}_1,\eta)\Phi(\bm{k}_2,\eta)\Phi(\bm{k}_3,\eta)\rangle
=(2\pi)^3\delta^D(\bm{k}_1+\bm{k}_2+\bm{k}_3)F_{\rm 2nd}(k_1,k_2,k_3,\eta),
\end{equation}
where 
\begin{equation}
 F_{\rm 2nd}(k_1,k_2,k_3,\eta)
=\frac{\eta^2}3\left[
\left(\frac{k_1k_2}{k_3}\right)^2F_2^{(s)}(\bm{k}_1,\bm{k}_2)
P_\Phi(k_1)P_\Phi(k_2)+(2~{\rm perm}.)
\right],
\label{eq:2nd}
\end{equation}
and $P_\Phi(k)=AT^2(k)/k^{4-n_s}$.
The shape dependence of $F_{\rm 2nd}(k_1,k_2,k_3)(k_2/k_1)^2(k_3/k_1)^2$
is shown in 
Figure~\ref{fig:2nd} for various values of $k_1$ (because $F_{\rm 2nd}$ is
not scale invariant). 
The CMB data are sensitive to $k_1<k_{\rm
 max}\sim 0.2~h~{\rm Mpc}^{-1}(l_{\rm max}/2000)$.
 We find that the bispectrum peaks at the equilateral configuration on
 large scales ($k_1<10^{-2}~h~{\rm Mpc}^{-1}$), and it peaks along the
 elongated configurations on small scales ($k_1\sim 0.1~h~{\rm
 Mpc}^{-1}$). It peaks at the squeezed configuration on a very small
 scale ($k_1\sim 1~h~{\rm  Mpc}^{-1}$), but these scales are not
 accessible by the CMB due to the Silk damping. 
 Note that the most squeezed configuration shown in this Figure has
 $k_1=k_2=100k_3$. The dominant shape changes with scales, as the linear
 transfer function, $T(k)$, declines with $k$, with the small-scale
 limit given by $T(k)\propto \ln k/k^2$.
 From these results, we
 expect the second-order effect in the Newtonian limit to yield only a small
 contamination of $\fnlKS$. 

The dominant contribution to the second-order temperature anisotropy in
the sub-horizon limit is
given by the second-order Sachs-Wolfe effect
\cite{pitrou/uzan/bernardeau:2008}:\footnote{
The CMB bispectrum from the second-order ISW effect was considered in
\cite{spergel/goldberg:1999,boubekeur/etal:2009}. 
}
\begin{equation}
 \frac{\Delta T^{(2)}}{T}(\hat{\bm{n}})
= \frac12R_*\Phi^{(2)}(r_*,\hat{\bm{n}}r_*),
\label{eq:sachs}
\end{equation}
where $R_*\equiv 3\rho_b/(4\rho_\gamma)$ is the baryon-photon ratio at
the decoupling epoch.
(Here, a factor of $1/2$ comes from our way of defining the second-order
temperature anisotropy, $\Delta T=\Delta T^{(1)}+\Delta T^{(2)}$ and the
second-order curvature perturbation,
$\Phi=\Phi^{(1)}+\frac12\Phi^{(2)}$. This definition follows from
Ref.~\cite{nitta/etal:2009}.) The corresponding second-order $a_{lm}$ is 
\begin{eqnarray}
\nonumber
 a_{lm}^{(2)}&=&4\pi(-i)^l
\int
\frac{d^3\bm{k}}{(2\pi)^3}\left[\frac12R_*\Phi^{(2)}(\bm{k},\eta_*)\right]j_l(kr_*)Y_{lm}^*(\hat{\bm{k}})\\
\nonumber
&=&
\frac{4\pi}8(-i)^l
\int
\frac{d^3\bm{k}}{(2\pi)^3}
\int\frac{d^3\bm{k}'}{(2\pi)^3}d^3\bm{k}''
\delta^D(\bm{k}'+\bm{k}''-\bm{k})
\Phi^{(1)}_p(\bm{k}')\Phi_p^{(1)}(\bm{k}'')
\\
\nonumber
&\times&\sum_{l'm'}\left[
\frac43R_*\left(\frac{k'k''\eta_*}{k}\right)^2
F_2^{(s)}(\bm{k}',\bm{k}'')T(k')T(k'')j_l(kr_*)\delta_{ll'}\delta_{mm'}\right]Y_{l'm'}^*(\hat{\bm{k}}),\\
\end{eqnarray}
where the linear primordial perturbation, $\Phi^{(1)}_p$, is related to $\Phi^{(1)}$ as
$\Phi^{(1)}(\bm{k})=\Phi^{(1)}_p(\bm{k})T(k)$. 
Comparing this with equation~\ref{eq:nitta}, we identify the term inside
the square bracket as the second-order radiation transfer function,
$F_{lm}^{l'm'}$.\footnote{Incidentally, in the notation of
\cite{nitta/etal:2009} (see 
their equation 2.29), ${\cal
S}_{00}^{(2)}(\bm{k}',\bm{k}'',\bm{k},\eta_*)=\frac43R_*\left(\frac{k'k''\eta_*}{k}\right)^2  
F_2^{(s)}(\bm{k}',\bm{k}'')T(k')T(k'')$.} 

With this result, one can calculate the reduced bispectrum of the Newtonian
second-order effect, $b_{l_1l_2l_3}^{\rm 2nd}$.
The resulting $\fnlKS$ is always
less than unity regardless of the angular scales (D. Nitta 2010, private
communication; also see
\cite{bartolo/riotto:2009,boubekeur/etal:2009}). The calculation was 
done for $l\le 2000$. 

How can we reconcile this result with those found in
\cite{pitrou/uzan/bernardeau:prep}? 
The calculations given above (equations~\ref{eq:newton}
and \ref{eq:sachs}) are valid only in the
sub-horizon limits, 
and thus they are not suitable for calculating the contribution to the
squeezed-limit bispectrum, which can correlate super- and sub-horizon
fluctuations. 
Therefore, the difference between these results
seems to imply:
\begin{itemize}
 \item [1.] The dominant contamination of $\fnlKS$ comes
from the general relativistic (post Newtonian) evolution of $\Phi^{(2)}$
that is not captured by the above Newtonian calculation
       (equation~\ref{eq:newton}). 
 \item [2.] The full second-order radiation transfer function beyond the
       sub-horizon approximation (equation~\ref{eq:sachs}) gives the
       dominant contribution to $\fnlKS$.
\end{itemize}
Perhaps both contributions are important.
 This is yet to
be confirmed; however, if this is true, one should be able to construct a simple
template for the second-order bispectrum, and use it to remove the
contamination by including its amplitude, $f_{\rm NL}^{\rm 2nd}$, in the
fit. 

\section{Four-point function: local-form trispectrum test of multi-field
 models}
Widely used notation for the ``local-form trispectrum'' is
\begin{eqnarray}
\nonumber
& & \langle
\Phi(\bm{k}_1)\Phi(\bm{k}_2)\Phi(\bm{k}_3)\Phi(\bm{k}_4)\rangle\\
\nonumber
&=& (2\pi)^3\delta^D(\bm{k}_1+\bm{k}_2+\bm{k}_3+\bm{k}_4)\\
\nonumber
&\times&\Bigl\{
\frac{25}{18}\tau_{\rm NL}\left[P_\Phi(k_1)P_\Phi(k_2)
\left\{P_\Phi(k_{13})+P_\Phi(k_{14})\right\}+(11~{\rm perm}.)\right]\\
& &+6g_{\rm NL}
\left[P_\Phi(k_1)P_\Phi(k_2)P_\Phi(k_3)+(3~{\rm perm}.)\right]
\Bigr\},
\end{eqnarray}
where $k_{ij}\equiv |\bm{k}_i+\bm{k}_j|$. (In this section, we use $\Phi$
for the primordial perturbation, i.e., $\Phi=\Phi_p=\frac35\zeta$.)
When the curvature perturbation is given by the simplest local form,
$\Phi=\Phi_L+\fnlKS\Phi_L^2+g_{\rm NL}\Phi_L^3$, one finds the above 
trispectrum with $\tau_{NL}=(6\fnlKS/5)^2$ \cite{boubekeur/lyth:2006}.
However, in general $\tau_{\rm NL}$ is different from
$(6\fnlKS/5)^2$. 

To see this, let us consider 
a broad class of multi-field models
in which the primordial curvature perturbation, $\zeta$, is given in
terms of the field derivatives of the number of $e$-folds, $N=\ln a$, and the 
perturbation in the $I$-th scalar field, $\delta\phi_I$, as
\begin{equation}
 \zeta=\sum_I\frac{\partial N}{\partial\phi_I}
\delta\phi_I+\frac12\sum_{IJ}\frac{\partial^2 N}{\partial\phi_I\partial\phi_J}\delta\phi_I\delta\phi_J+\dots,
\end{equation}
where $\langle\delta\phi_I\delta\phi_J\rangle=0$ for $I\neq
J$. This expansion is known as the ``$\delta N$ formalism''
\cite{starobinsky:1982,starobinsky:1985b,sasaki/stewart:1996,salopek/bond:1990,lyth/rodriguez:2005}.
In this case, $\fnlKS$ and $\tau_{\rm NL}$ are given by
\cite{lyth/rodriguez:2005,alabidi/lyth:2006b,byrnes/sasaki/wands:2006}
\begin{eqnarray}
 \frac65\fnlKS&=&
  \frac{\sum_{IJ}N_{,IJ}N_{,I}N_{,J}}{[\sum_{I}(N_{,I})^2]^2},\\
\tau_{\rm NL}&=&
\frac{\sum_{IJK}N_{,IJ}N_{,J}N_{,IK}N_{,K}}{[\sum_{I}(N_{,I})^2]^3}=
\frac{\sum_{I}(\sum_{J}N_{,IJ}N_{,J})^2}{[\sum_{I}(N_{,I})^2]^3},
\end{eqnarray}
where $N_{,I}\equiv \partial N/\partial\phi_I$ and 
$N_{,IJ}\equiv \partial^2 N/\partial\phi_I\partial\phi_J$.
Suyama and Yamaguchi \cite{suyama/yamaguchi:2008} showed that
the Cauchy-Schwarz inequality implies that 
the following
inequality,
\begin{equation}
 \tau_{\rm NL}\ge \left(\frac{6\fnlKS}{5}\right)^2,
\label{eq:suyama}
\end{equation}
is satisfied. To derive this result, use the Cauchy-Schwarz inequality:
\begin{equation}
 \left(\sum_Ia_I^2\right)\left(\sum_Jb_J^2\right)\ge
  \left(\sum_Ia_Ib_I\right)^2, 
\end{equation}
with 
\begin{eqnarray}
 a_I&=& \frac{\sum_JN_{,IJ}N_{,J}}{[\sum_J(N_{,J})^2]^{3/2}},\\
 b_I&=& \frac{N_{,I}}{[\sum_J(N_{,J})^2]^{1/2}}.
\end{eqnarray}
The equality, $\tau_{\rm NL}=({6\fnlKS}/{5})^2$ is
satisfied for the simplest local-form model,
$\zeta=N_1\delta\phi_1+(N_{11}/2)\delta\phi_1^2$.

Note, however, that one finds a different relation between $\fnlKS$ and
$\tau_{\rm NL}$ when the Cauchy-Schwarz inequality becomes trivial,
i.e., $0=0$. For example, when $\zeta$ is given by \cite{boubekeur/lyth:2006}
\begin{equation}
 \zeta=\frac{\partial N}{\partial\phi_1}
\delta\phi_1+\frac12\frac{\partial^2 N}{\partial\phi_2^2}\delta\phi_2^2,
\label{eq:lyth}
\end{equation}
and $\langle\delta\phi_1\delta\phi_2\rangle=0$, 
one finds $\tau_{\rm NL}\sim 10^3(\fnlKS)^{4/3}$
\cite{suyama/takahashi:2008}. The Cauchy-Schwarz inequality becomes
$0=0$ because $a_I=0$ for all $I$, and thus both the bispectrum and the
trispectrum come from the second term in equation~\ref{eq:lyth}: the
bispectrum is given by the 6-point function of $\delta\phi_2$, and the
trispectrum is given by the 8-point function of $\delta\phi_2$. 

In this case, whether
$\tau_{\rm NL}\ge (6\fnlKS/5)^2$ is satisfied depends on the value of
$\fnlKS$. For this particular example, the current limit of $\fnlKS<74$
implies that $\tau_{\rm NL}\ge (6\fnlKS/5)^2$ is still satisfied
parametrically. 

The inequality is valid also for the ``quasi-single field
inflation'' of \cite{chen/wang:2010}; see equation~(7.6) of
\cite{chen/wang:prepb}. (Note that $\tau_{NL}^{\rm SE}$ and 
$\tau_{NL}^{\rm CI}$ in \cite{chen/wang:prepb} correspond
to $\tau_{\rm NL}$ and $g_{\rm NL}$ in this paper, respectively.)
 
Therefore, if observations indicate $\tau_{\rm
NL}<(6\fnlKS/5)^2$, a broad class of multi-field
models satisfying the above conditions would be ruled out.
This property makes the trispectrum a
powerful probe of the physics of multi-field models.
Note that the simplest local-form limit, $\tau_{\rm
NL}=(6\fnlKS/5)^2$, has no special significance, as this is just one of
many possibilities of multi-field models. (All single-field models
predict a non-detectable level of the primordial $\fnlKS$, and thus the
observational test using the above relation between $\fnlKS$ and $\tau_{\rm
NL}$ has no relevance to single-field models.)

The expected 95\% uncertainties in $\tau_{\rm NL}$ from the 
7-year {\sl WMAP} data ($l_{\rm max}\sim 500$) and 
the 1-year {\sl Planck} data ($l_{\rm max}\sim 1500$) are 
5000 and 560, respectively \cite{kogo/komatsu:2006}.
{\it If the {\sl Planck} finds $\fnlKS\sim 30$, then it would be able to
test if the measured $\tau_{\rm NL}$ would satisfy equation~\ref{eq:suyama}.}
This provides an excellent science case for the trispectrum that would
be measured by {\sl Planck}.

The expected uncertainties in $g_{\rm NL}$ have not been calculated yet,
although we expect them to be much greater than those for
$\tau_{\rm NL}$, as $g_{\rm NL}$ is the coefficient of the cubic-order
term (i.e., $g_{\rm NL}$ is much more difficult to constrain than
$\tau_{\rm NL}$) \cite{creminelli/senatore/zaldarriaga:2007}. 

The local-form trispectrum is not the only possibility. 
Various other inflation models 
predict distinctly different quadrilateral shape dependence.
For some general analyses of shapes, see 
\cite{chen/etal:2009,gao/li/lin:2009,renaux-petel:2009,chen/wang:prepb}.

Finally, while we do not discuss the large-scale structure of the
universe in this article, the most promising probe of the local-form
trispectrum seems to be the {\it bispectrum} of galaxies.
See \cite{jeong/komatsu:2009,sefusatti:2009,giannantonio/porciani:prep}
for details. 

\section{Current Results}

\subsection{Bispectrum}
(Most of this subsection is adopted from Section~6.2 of
\cite{komatsu/etal:prep}.) 
\begin{table}[t]
\centering
\begin{tabular}{lccccc}
\hline
Band
& Foreground
& $\fnlKS$
& $\fnleq$
& $\fnlor$
& $b_{\rm src}$\\
\hline
V+W & Raw  &  $59\pm 21$  &  $33\pm 140$  &  $-199 \pm 104$  & N/A \\
V+W & Clean & $42\pm 21$  &  $29\pm 140$  &  $-198 \pm 104$  & N/A \\
V+W & Marg. & $32\pm 21$  &  $26\pm 140$  &  $-202\pm
 104$ & $-0.08\pm 0.12$\\
V   & Marg. & $43\pm 24$  &  $64\pm 150$  &  $-98 \pm 115$ & $0.32\pm
 0.23$ \\
W   & Marg. &  $39\pm 24$  &  $36\pm 154$  &  $-257\pm 117$  & $-0.13\pm
 0.19$\\
\hline
\label{tab:fnl}
\end{tabular}
\caption{Estimates and the corresponding 68\% intervals of the
 primordial non-Gaussianity parameters ($\fnlKS$, $\fnleq$, $\fnlor$)
 and the point source bispectrum amplitude, $b_{\rm src}$ (in units of
 $10^{-5}~\mu{\rm K}^3~{\rm sr}^2$), from the {\sl WMAP} 7-year
 temperature maps. This table is adopted from \cite{komatsu/etal:prep}.} 
\end{table}

In 2002, the first limit on $\fnlKS$ was obtained from the {\sl COBE}
4-year data 
\citep{bennett/etal:1996} by \cite{komatsu/etal:2002}, using the
angular bispectrum. The limit was
improved by an order of magnitude when the {\sl WMAP} first year data were
used to constrain $\fnlKS$ \citep{komatsu/etal:2003}. Since then the
limits have improved steadily as {\sl WMAP} collects more years of data and
the bispectrum method for estimating $\fnlKS$ has improved
\citep{komatsu/spergel/wandelt:2005,creminelli/etal:2006,creminelli/etal:2007,spergel/etal:2007,yadav/wandelt:2008,komatsu/etal:2009,munshi/heavens:2010,smidt/etal:2009,smith/senatore/zaldarriaga:2009}.  

Using the optimal estimators described in Section~3, we have constrained
the primordial non-Gaussianity parameters as well as the point-source
bispectrum using the {\sl WMAP} 7-year data.
The 7-year data and results are described in
Refs.~\cite{larson/etal:prep,komatsu/etal:prep,jarosik/etal:prep,gold/etal:prep,bennett/etal:prep,weiland/etal:prep}. 
 
We use the V- and W-band maps at the HEALPix resolution $N_{\rm
side}=1024$. As the optimal estimator weights the data optimally at all
multipoles, we no longer need to choose the maximum multipole used in
the analysis, i.e., we use all the data. We use both the
raw maps (before cleaning foreground) and foreground-reduced (clean)
maps to quantify the  foreground contamination of $f_{\rm NL}$ parameters.
For all cases,
we find the best limits on $f_{\rm NL}$ parameters by combining the V-
and W-band maps, and marginalizing
over the synchrotron, 
free-free, and dust foreground templates \citep{gold/etal:prep}.
As for the mask, we always use the {\it KQ75y7} mask \citep{gold/etal:prep}.

In Table~\ref{tab:fnl}, we summarize our results:
\begin{itemize}
 \item [1.] {\bf Local form results}. The 7-year best estimate of $\fnlKS$ is
$$\fnlKS=32\pm 21~\mbox{(68\%~CL)}.$$
The 95\% limit is $-10<\fnlKS<74$. When the raw maps are used, we find 
$\fnlKS=59\pm 21$~(68\%~CL). When the clean maps are used, but
       foreground templates are not marginalized over, we find 
$\fnlKS=42\pm 21$~(68\%~CL). These results (in particular the clean-map
       versus the foreground marginalized) indicate that the foreground
       emission makes a difference at the level of $\Delta\fnlKS~\sim
       10$.\footnote{The effect of the foreground marginalization
       depends on an estimator. Using the 
       needlet bispectrum, Cabella et al. \cite{cabella/etal:prep} found
       $\fnlKS=35\pm 42$ and $38\pm 47$~(68\%~CL) with and without the
       foreground 
       marginalization, respectively.} We find that the V+W result is lower than the V-band or W-band
       results. This is possible, as the V+W result contains
       contributions from the 
       cross-correlations of V and W such as $\langle {\rm VVW}\rangle$ and
       $\langle {\rm VWW}\rangle$. 
\item [2.] {\bf Equilateral form results}. The 7-year best estimate of $\fnleq$ is
$$\fnleq=26\pm 140~\mbox{(68\%~CL)}.$$
The 95\% limit is $-214<\fnleq<266$. For $\fnleq$, the foreground
      marginalization does not shift the central values very much,
      $\Delta\fnleq=-3$. This makes sense, as the equilateral bispectrum
      does not couple small-scale modes to very large-scale modes
      $l< 10$, which are sensitive to the
      foreground emission. On the other hand, the local form bispectrum
      is dominated by the squeezed triangles, which do couple large and
      small scales modes.
\item [3.] {\bf Orthogonal form results}. The 7-year best estimate of $\fnlor$ is
$$\fnlor=-202\pm 104~\mbox{(68\%~CL)}.$$
The 95\% limit is $-410<\fnlor<6$. The foreground marginalization has
little effect, $\Delta\fnlor=-4$.
\end{itemize}

As for the point-source bispectrum, we do not detect $\bsrc$ in V, W, or
V+W. In \cite{komatsu/etal:2009}, we estimated that the residual
sources could bias $\fnlKS$ by a small positive amount, and applied corrections
using Monte Carlo simulations. In this paper, we do not attempt to make
such corrections, but we note that sources could give $\Delta \fnlKS\sim
2$ (note that the simulations used by \cite{komatsu/etal:2009} likely
overestimated the effect of sources by a factor of two). As the
estimator has changed from that used by \cite{komatsu/etal:2009},
extrapolating the previous results is not trivial. Source
corrections to $\fnleq$ and $\fnlor$ could be larger
\cite{komatsu/etal:2009}, but we have not estimated the magnitude
of the effect for the 7-year data.

As we described in Section~4, among various sources of secondary non-Gaussianities which might
contaminate measurements of primordial non-Gaussianity (in
particular $\fnlKS$), a 
coupling between 
the ISW effect and the weak gravitational lensing is the most dominant
source of confusion for $\fnlKS$. 
Calabrese et al. 
\cite{calabrese/etal:2010} used the skewness power spectrum method of \cite{munshi/etal:prep} to 
search for this term in the {\sl WMAP} 5-year data and found a null
result.

\subsection{Trispectrum}
The optimal estimators for the trispectrum have not been implemented,
largely because they are computationally demanding. 
While the first measurements of the angular bispectrum were made from the 
the {\sl COBE} 4-year data \citep{bennett/etal:1996}
by \cite{komatsu:prep,kunz/etal:2001} in 2001, limits on the physical
parameters have not been obtained from the direct trispectrum analysis.

Recently, Smidt et al. \cite{smidt/etal:prep} used the sub-optimal
estimator developed in 
\cite{munshi/etal:prepb} (which becomes optimal in the limit that the
instrumental noise
is isotropic) and found
the 95\%~CL limits of 
$-7.4\times 10^5<g_{\rm NL}<8.2\times 10^5$ and
$-0.6\times 10^4<\tau_{\rm NL}<3.3\times 10^4$. 
The current limit is consistent
with the Suyama-Yamaguchi inequality, $\tau_{\rm NL}\ge (6\fnlKS/5)^2$.

\subsection{Other statistical methods}
While the optimal estimators for the $f_{\rm NL}$ parameters (with the
minimum variance) must be constructed from the PDF as in Section~3,
there are various other ways of constraining
non-Gaussianity. While these other methods are usually sub-optimal, they
serve as useful diagnosis tools of the results obtained from the direct
bispectrum and trispectrum methods. In some cases, they are easier to
implement than the optimal estimators. 

A major progress in the topological Gaussianity test using the {\it Minkowski
functionals}
\cite{gott/etal:1990,mecke/buchert/wagner:1994,schmalzing/buchert:1997,schmalzing/gorski:1998,winitzki/kosowsky:1998}
since 2004 is the derivation and implementation of the analytical
formula for the Minkowski functionals of the CMB
\cite{hikage/komatsu/matsubara:2006,matsubara:2010}. This method has
been applied to the {\sl WMAP} data
\cite{hikage/etal:2008,hikage/etal:2009} as well as to the BOOMERanG
data \cite{natoli/etal:2009}. The {\sl Planck} data are expected to
reach the 68\% limit of $\Delta\fnlKS=20$
\cite{hikage/komatsu/matsubara:2006}, which is worse than the limit 
from the optimal method, $\Delta\fnlKS=5$
\cite{komatsu/spergel:2001}. An advantage of the Minkowski functionals
is that the measurements of the
Minkowski functionals do not depend on the models, and thus the
computational cost is the same for all models. This allows one to obtain
limits on various models, for which the optimal estimators are difficult
to implement. For example, a limit on the primordial non-Gaussianity in
the isocurvature perturbation is currently available only from 
the Minkowski functionals \cite{hikage/etal:2009}. 

Instead of expanding the temperature anisotropy into spherical
harmonics, one may choose to expand it using a different basis. One
popular basis used in the CMB community is the so-called {\it Spherical
Mexican Hat Wavelet} (SMHW). See
\cite{martinez-gonzalez/etal:2002,martinez-gonzalez:prep} for reviews on
this method.
A major progress in this method is the realization that the 3-point
function of the wavelet coefficients made of large and small smoothing
scales is nearly an optimal estimator for the local-form
bispectrum: when only the adjacent scales are included, Curto et
al. \cite{curto/etal:2009} found $-8<\fnlKS<111$ (95\%~CL) from the {\sl
WMAP} 5-year data. When all the scales (including large-small scale
combinations) are included in the analysis, the limit improved
significantly to $-18<\fnlKS<80$ (95\%~CL)
\cite{curto/martinez-gonzalez/barreiro:2009}, which is similar to the
optimal limit from the 5-year data, $-4<\fnlKS<80$ (95\%~CL)
\cite{smith/senatore/zaldarriaga:2009}. 
An advantage of the SMHW is that it retains information on the spatial
distribution of the signal. This property can be used to measure
$\fnlKS$ as a function of positions on the sky
\cite{curto/martinez-gonzalez/barreiro:2009}. 
In addition, the analytical formula for the SMHW 
as a function of $\fnlKS$ has been derived and
implemented (A. Curto 2009, private communication).
See \cite{mukherjee/wang:2004,cabella/etal:2005} for earlier limits on
$\fnlKS$ from the SMHW. 

Another form of spherical wavelets that has been used to constrain
$\fnlKS$ is the {\it spherical needlets} \cite{marinucci/etal:2008}.
The limits on $\fnlKS$ are reported in 
\cite{rudjord/etal:2009,pietrobon/etal:2009,cabella/etal:prep}.
This method also allows one to look for a spatial variation in $\fnlKS$,
and the results are reported in
\cite{rudjord/etal:2010,pietrobon/etal:2010}.  
For the other types of wavelets considered in the literature, see
\cite{aghanim/etal:2003,jin/etal:2005} and references therein.

Many other statistical methods have been proposed and used for
constraining $\fnlKS$ in the 
literature. An incomplete list of references is:
\cite{gaztanaga/wagg:2003,chen/szapudi:2005} on the real-space 3-point
function; \cite{cabella/etal:2006} on the integrated bispectrum;
\cite{chen/szapudi:2006} on the 2-1 cumulant correlator;
\cite{cabella/etal:2005} on the local curvature; and 
\cite{vielva/sanz:2009,vielva/sanz:prep} on the $N$-point PDF.
Also see references therein. While we have not listed the statistical methods
that have not been used to constrain the primordial non-Gaussianity
parameters yet, there are many other methods proposed in a general
context in the literature.

\section{Conclusion}
Since the last review articles on signatures of primordial
non-Gaussianity in the CMB were written in 2001 \cite{komatsu:prep}
and 2004 \cite{bartolo/etal:2004}, a lot of progress has been made in this
field. The current standard lore may be summarized as follows:
\begin{itemize}
 \item[1.] {\bf Shape and physics.} Different aspects of the physics of
	   the primordial universe 
	   appear in different shapes of three- and four-point
	   functions.
 \item[2.] {\bf Importance of local shape.} Of these shapes, the local
	   shapes have special significance: 
	   a significant detection of the local-form bispectrum (with
	   $\fnlKS\gg 1$) would rule out {\it all} single-field
	   inflation models, and the local-form trispectrum can be used
	   to rule out 
	   a broad class 
	   of multi-field models (if not {\it all} multi-field models)
	   by testing $\tau_{\rm NL}\ge 
	   (6\fnlKS/5)^2$.
 \item[3.] {\bf Optimal estimators.} The optimal estimators of the
	   bispectrum and trispectrum can 
	   be derived systematically from the expansion of the PDF. The
	   optimal bispectrum estimator has been implemented. 
 \item[4.] {\bf Secondary.} The most serious contamination of $\fnlKS$
	   is due to the 
	   lensing-ISW coupling, which can be removed by using the
	   template given in Section~4.
 \item[5.] {\bf Foreground.} The Galactic foreground contamination is
	   minimal for $\fnleq$ 
	   and $\fnlor$, but it can be as large as $\fnlKS\sim 10$ for
	   the local-form bispectrum. This
	   must be carefully studied and eliminated in the {\sl Planck}
	   data analysis. The random (Poisson) point-source
	   contamination can be removed by using the template given in
	   Section~3.2. 
\end{itemize}
Some outstanding issues for the ``CMB and primordial non-Gaussianity'' include:
\begin{itemize}
 \item[1.] {\bf Second order.} (In Newtonian gauge) the products of the
	   first-order terms and the intrinsically second-order terms in
	   the sub-horizon limit do not contaminate the local-form
	   bispectrum very much ($\Delta\fnlKS<1$). However, would the
	   post-Newtonian effect give $\Delta\fnlKS\sim 5$, as found by 
	   \cite{pitrou/uzan/bernardeau:prep}? If so, we need to
	   construct a template for this effect.
 \item[2.] {\bf More foreground.} How can we model the non-Poisson
	   (clustered) point source bispectrum? How about the foreground
	   (and secondary) contamination of the primordial trispectrum?
 \item[3.] {\bf Trispectrum estimators.} How can we implement the optimal
	   trispectrum estimators for both local and non-local shapes?
\end{itemize}
These issues would become important when the {\sl Planck} data are
analyzed in search of primordial non-Gaussianity.
The {\sl Planck} is expected to reduce the uncertainty in $\fnlKS$ by a
factor of four compared to the current limit, $\fnlKS=32\pm 21$
(68\%~CL). If the {\sl Planck} detected $\fnlKS\sim 30$, then the
trispectrum would provide an important test of multi-field models. 
In particular, if $\fnlKS\gg 1$ and $\tau_{\rm NL}<(6\fnlKS/5)^2$ are
found, then {\it all} single-field models and many (if not {\it all})
multi-field models would be ruled out, and thus the standard paradigm
of inflation as the origin of fluctuations would face a serious challenge.

However, do not despair even if the {\sl Planck} did not detect the
primordial bispectrum or trispectrum - while the CMB may end its leading
role as a probe of primordial non-Gaussianity (unless the
next-generation, comprehensive CMB
satellite which can measure both the temperature and polarization to the
cosmic-variance-limited precision is funded \cite{baumann/etal:2009}),
the large-scale  
structure of the universe would eventually take over and substantially
reduce the uncertainties in the local-form parameters such as $\fnlKS$,
$\tau_{\rm NL}$, and $g_{\rm NL}$ 
(see Desjacques and Seljak's article in this volume).

\section*{Acknowledgments} 
We thank the organizers and participants of ``The Non-Gaussian
 Universe'' workshop at Yukawa Institute for Theoretical Physics (YITP)
 for stimulating discussion and presentations, D. Jeong for making
 Figure~\ref{fig:prim} and \ref{fig:2nd}, D. Nitta for
 calculating $\fnlKS$ from the second-order $\Phi$
 in the Newtonian limit, and T. Takahashi for valuable comments on
 Section~6. This work is supported in part by  
an NSF grant PHY-0758153.
\providecommand{\newblock}{}

\end{document}